\pdfoutput=1
\documentclass[nofootinbib,twocolumn, prd, preprintnumbers]{revtex4-1}
\usepackage{amsmath,amssymb,graphicx,bm,psfrag,color,slashed,subcaption,array,subfiles,cancel}
\usepackage[version=4]{mhchem}
\usepackage[colorlinks=true, allcolors=blue]{hyperref}
\usepackage[dvipsnames]{xcolor}

\usepackage{siunitx} 
\DeclareSIUnit{\year}{y}

\definecolor{rev1}{rgb}{1, 0, 0}

\begin{document}

\preprint{IPPP/23/45}
\vspace{0.5cm}
\title{\boldmath
How to measure the spin of invisible states in \texorpdfstring{$e^+e^- \to \gamma + X$}{eeaX}}
\author{Martin Bauer}
\author{Sofie Nordahl Erner}

\address{Institute for Particle Physics Phenomenology, Department of Physics\\
Durham University, Durham, DH1 3LE, United Kingdom}

\begin{abstract}
We examine the production of an invisible state $X$ together with a photon, $e^+e^- \to \gamma +X$, at electron positron colliders and present measurement strategies that can detect the spin of the invisible state as well as the underlying production mechanism, based on the angular distribution of the final state photon, the cross-sections for polarized initial states, and the photon polarization. Our measurement strategy can be used to identify whether the invisible state is a hidden photon or an axion. The results are compared with a detailed analysis of the Standard Model background, and we calculate the sensitivity reach for searches for axions and hidden photons at Belle II.
\end{abstract}

\maketitle

%%%%%%%%%%%%%%%%%%%%%%%%%%%%%%%%%
%%%%%%%%%%%%%%%%%%%%%%%%%%%%%%%%%
\section{Introduction} \label{sec:intro}
%%%%%%%%%%%%%%%%%%%%%%%%%%%%%%%%%
%%%%%%%%%%%%%%%%%%%%%%%%%%%%%%%%%
Searches for missing energy at electron positron colliders with a final state photon, $e^+e^- \to \gamma +X$, are one of the most sensitive probes for invisible states $X$ with masses in the range $~ 0.1-9$ GeV. With unequal beam energies, $E_{e^+} = 4~\text{GeV}$ and $E_{e^-} = 7~\text{GeV}$, and a centre-of-mass energy of $\sqrt{s} = 10.58~\text{GeV}$, Belle II will provide a valuable opportunity to search for these signatures ~\cite{Belle-II:2018jsg}.
Here we discuss how we can take advantage of the clean environment at $e^+e^-$ colliders and use the angular distribution of the final state photons, polarized beams, and final state photon polarization to identify the spin and the coupling structure of the invisible state. We introduce the most minimal model of a spin 1 hidden photon (also known as a dark photon) interacting with electrons, and a pseudoscalar spin 0 boson interacting with electrons and photons, which we call axions throughout the paper.
For axions such a search has been proposed first by Wilczek~\cite{Wilczek:1977zn} and searches have been performed  with CLEO~\cite{CLEO:1994hzy} and BaBar~\cite{BaBar:2008aby,BaBar:2010eww}. For Belle II, projections are available~\cite{Dolan:2017osp, Darme:2020sjf, Belle-II:2022cgf}. A search for hidden photons has been performed by BaBar~\cite{BaBar:2017tiz} and a projection for Belle II can be found in~\cite{Belle-II:2018jsg}. 
The existing searches assume a model and production mechanism, and interpret searches for missing energy and a single photon as a signal in that model. We extend the search strategy proposed in~\cite{Belle-II:2018jsg} and identify observables that can distinguish between the different models in case a signal is observed.\\
First, we demonstrate that the angular distribution can be used to determine whether the invisible state has been produced in an s-channel or t/u-channel process. 
Second, we suggest extending the proposed 'Chiral Belle programme' that aims to use a 70\% polarized electron beam in a future upgrade of Belle II~\cite{Forti:2022mti}. 
We argue that a polarization of both the electron beam and the positron beam allows a measurement that would uniquely fix the quantum numbers of the invisible state. In addition, Standard Model background processes are largely suppressed and allow for more aggressive cuts on the photon energy and scattering angle. We discuss the different processes contributing to this background and propose new strategies that can lead to improved sensitivity. For the three models discussed, we compute the expected sensitivity reach at Belle II. 
An alternative, more speculative strategy is to explore the possibility to detect the polarization of the final state photon. We discuss how such a measurement would give additional, complementary information on the production process and quantum numbers of the invisible state.

The rest of the paper is organised as follows.
In Sec. \ref{sec:DM_lag}, we introduce the hidden photon and axion Lagrangians, and 
in Sec. \ref{sec:amp_form} we discuss how the differential cross-sections with and without polarized beams can be used to identify the production process.
In Sec. \ref{sec:sims}, simulations and analyses for signal and background are presented, and we propose to improve the search strategy. We discuss the implications of measuring the photon polarization in Sec. \ref{sec:photonpolarization} and
in Sec. \ref{sec:NP} we present the sensitivity reach for future runs of Belle II.

%%%%%%%%%%%%%%%%%%%%%%%%%%%%%%%%%
%%%%%%%%%%%%%%%%%%%%%%%%%%%%%%%%%
\section{Axion and hidden photon Lagrangian} \label{sec:DM_lag}
%%%%%%%%%%%%%%%%%%%%%%%%%%%%%%%%%
%%%%%%%%%%%%%%%%%%%%%%%%%%%%%%%%%
We compare two minimal models of invisible states that can be produced via $e^+e^-\to \gamma + X$ that carry either spin 1 or spin 0. The relevant terms in the canonically normalized basis for the spin 1 field or hidden photon are given by,
\begin{align}\label{eq:DP_lag}
    \mathcal{L} \ni -c_{X} \bar \psi \gamma_\mu \psi X^\mu - \frac{m_X^2}{2} X_\mu  X^\mu \,,
\end{align}
where $m_X$ is the mass of the hidden photon and $c_X$ denotes its coupling strength to electrons. The interaction can be proportional to a new gauge charge in the case $X$ is the gauge boson of a new gauge group under which one of the SM global symmetries is charged, or generated via mixing with the SM photon. In the case of kinetic mixing one can write $c_X= \epsilon\, e Q_\psi$, where $\epsilon$ is the coefficient of the kinetic mixing term of the electromagnetic field strength tensor $F_{\mu\nu}$ and the hidden photon field strength tensor, $\mathcal L \ni -\epsilon/2 \, F_{\mu\nu}X^{\mu\nu}$, and $Q_\psi$ is the electric charge of the fermion $\psi$ in units of the electron charge $e$~\cite{Holdom:1985ag, Foot:1990mn, He:1990pn, He:1991qd}. Interaction due to mixing with the Z boson are suppressed for masses $m_X\leq 10$ GeV, but can play a role for hidden photons with masses closer to the $Z$-pole~\cite{Bauer:2018onh,Bauer:2022nwt}.

In the case of the spin 0 particle we consider an axion $a$ with the interactions,
\begin{align}\label{eq:ALP_lag}
    \mathcal{L} \ni c_e\frac{\partial_\mu a}{2f} \bar \psi \gamma^\mu \gamma_5 \psi - c_{\gamma} \frac{a}{f}F_{\mu\nu} \tilde F^{\mu\nu}-\frac{m_a^2}{2}a^2\,,
\end{align}

where $c_e$ and $c_{\gamma}$ are coupling constants, and $m_a$ is the mass term, though we also use $m_X$ to denote the mass of the new particle when referring to both the hidden photon and axions. Pseudoscalars with these interaction terms are also called Axion-Like-Particles (ALPs) in the literature. Since any axion will generically interact with photons and leptons due to renormalization group running we use the term \emph{axion} here~\cite{Bauer:2020jbp}.

\begin{figure}[tbp]
    \centering
    \includegraphics[width=.49\textwidth]{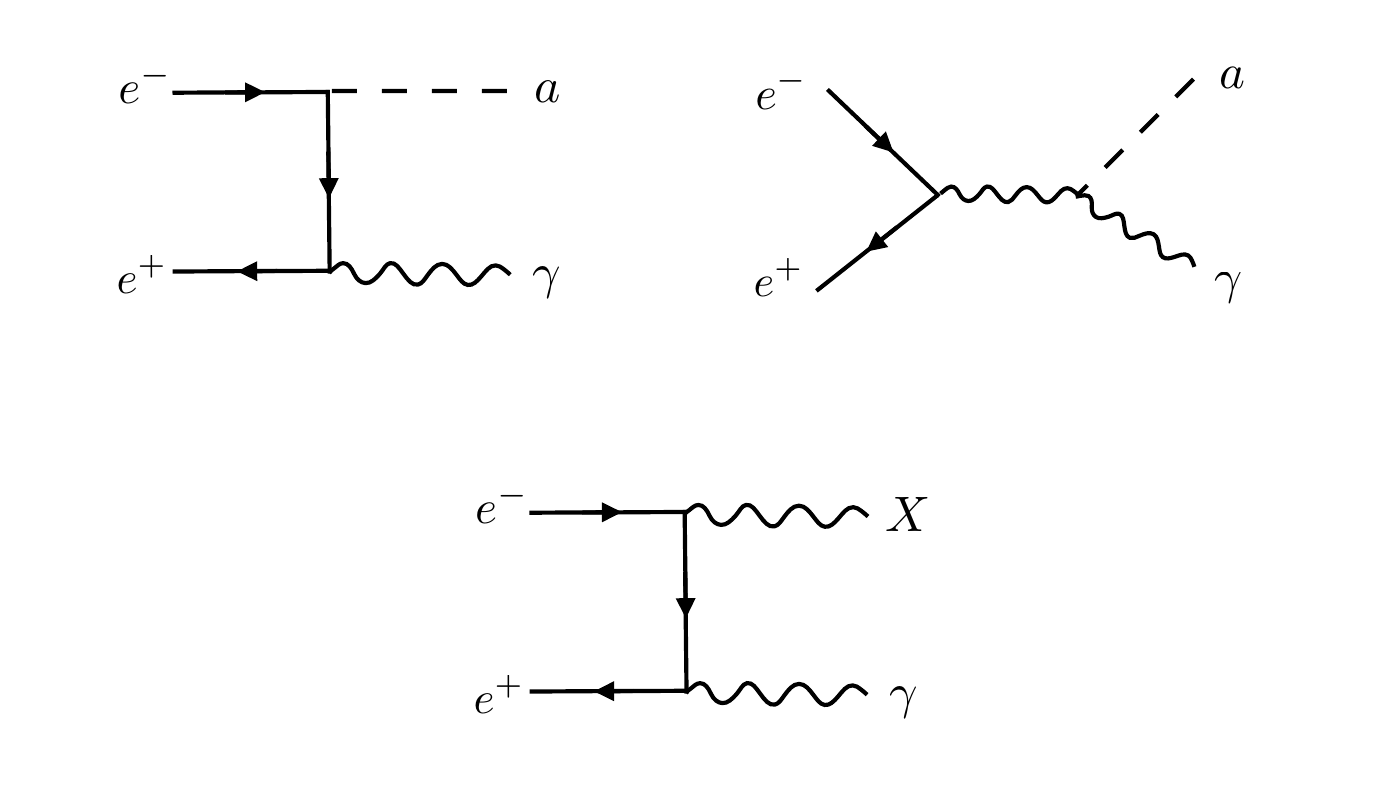}
    \caption{Different contributions to the production of axions (top) and hidden photons (bottom) at an $e^+e^-$ collider.}
    \label{fig:feyn}
\end{figure}

In all cases we consider the hidden photon or axion not to decay into SM particles on collider scales, e.g. by introducing a dominant decay into a set of invisible particles like dark matter. Otherwise for the mass range considered here with masses up to $m_X= 9$ GeV, both the hidden photon and axion decay back into electron positron or photon pairs. Neglecting subleading terms suppressed by the electron mass, the corresponding decay lengths for the three models read,
\begin{align}
\ell (c_X)&=\left(\frac{c_X^2m_X}{12\pi}\right)^{-1}\approx  10^{-12} \left[\frac{1}{m_X c_X^2}\right]\text{GeV}\text{m}\,,\notag\\
\ell (c_e)&=\left(\frac{c_e^2m_a m_e^2}{8\pi f^2}\right)^{-1}\approx 2\times 10^{-8} \left[\frac{f^2}{m_a c_e^2}\right]\text{GeV}^{-1}\text{m}\,,\notag\\
\ell (c_\gamma)&=\left(\frac{c_{\gamma}^2 m_a^3 }{4\pi f^2}\right)^{-1}\approx 2\times 10^{-15} \left[\frac{f^2}{m_a^3 c_{\gamma}^2}\right]\text{GeV} \text{m}\,.
\end{align}
The main differences between the models are their spin, the coupling structure to fermions, and the axion coupling to photons. The production of the hidden photon occurs only though couplings to the electron, whereas for the axion there are two contributions. The corresponding diagrams for are shown in Figure~\ref{fig:feyn}.

%%%%%%%%%%%%%%%%%%%%%%%%%%%%%%%%%
%%%%%%%%%%%%%%%%%%%%%%%%%%%%%%%%%
\section{Angular distributions} \label{sec:amp_form}
%%%%%%%%%%%%%%%%%%%%%%%%%%%%%%%%%
%%%%%%%%%%%%%%%%%%%%%%%%%%%%%%%%%

\begin{figure}[tbp]
    \centering
\includegraphics[width=.51\textwidth]{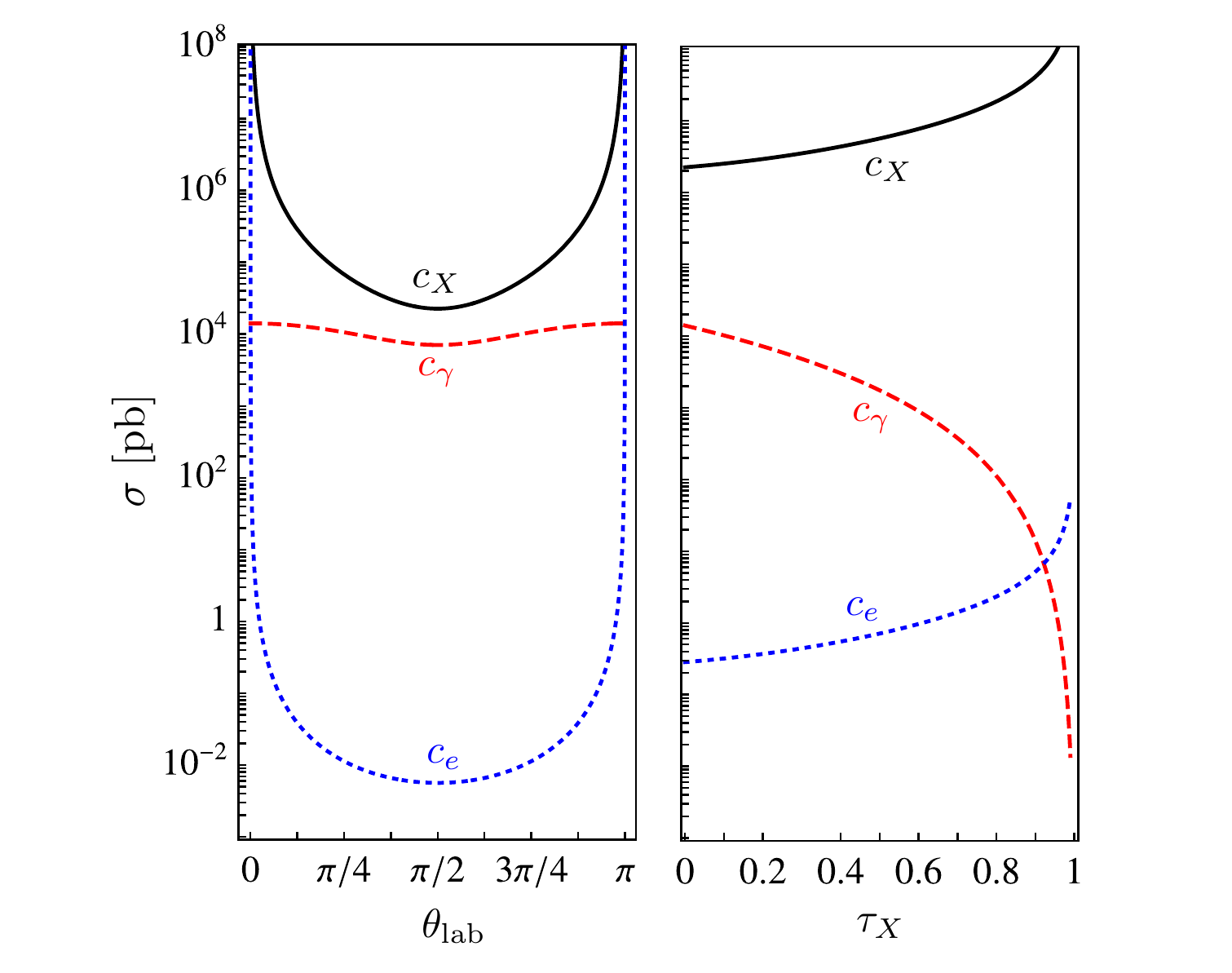}
    \caption{Cross-sections for the production of hidden photons with $c_X=1$ (black), axions with couplings to photons for $c_\gamma/f=1$/GeV and $c_e=0$ (red dashed), and electrons with $c_e/f=1$/GeV and $c_\gamma=0$ (blue dotted) as a function of the scattering angle $\theta_\text{lab}$ (left) and $\tau_X$ (right). We set $s=10$ GeV, $\tau_X=0$ (left), and $\theta_\text{lab}=\pi/4$ (right).}
    \label{fig:diffcs}
\end{figure}

In the following we discuss how the angular distribution of the final state photon in the process $e^+e^-\to \gamma+X$ can be used to discriminate between the hidden photon and axion final states as well as between the axion coupling to electrons and photons. 
In general the amplitude for the process,
\begin{equation}\label{eq:proc}
    e^+(p_1,\lambda_{e^+}) + e^-(p_2,\lambda_{e^-}) \to \gamma(q_1,\lambda_\gamma) + X(q_2,\lambda_X)\,,
\end{equation}
can be written as $\mathcal{M} = \mathcal{M}_\mu \epsilon^\mu(q_1,\lambda_\gamma)$ with the photon polarization vector $\epsilon^\mu$ and helicities 
$|\lambda_{e^\pm}| = 1/2$, $|\lambda_\gamma| = 1$, $|\lambda_X| = 0$ for the axion, and $|\lambda_X| = 1$ for the hidden photon, respectively. 
At Belle II, the incoming beams are angled $83 $ mrad with respect to each other, with the z-axis defined with equal distance to the beams \cite{Belle-II:2018jsg}.
This distinction is assumed to have little effect on the results presented in this paper, and hence it will be assumed that the beams are antiparallel along the z-axis.
We define the momenta of the incoming particles parallel to the $z$-axis.
The differential cross-section for the production of hidden photons in the centre-of-mass frame reads up to corrections of order $m_e^2/s$,
\begin{align}
\frac{d\sigma}{d\Omega}=c_X^2  \frac{\alpha}{4\pi s}\frac{(1+\tau_X)^2+(1-\tau_X)^2\cos^2\theta}{(1-\tau_X)(1-\cos^2\theta)} \,,
\end{align}
where $s$ is the centre-of-mass energy squared, $\cos\theta$ is the angle between the photon and the beam axis in the lab frame, and $\tau_X=m_X^2/s$. For the production of axions the differential cross-section reads~\cite{Darme:2020sjf},
\begin{align}\label{eq:axiondiffcs}
\frac{d\sigma}{d\Omega}&=\frac{\alpha}{4\pi f^2}\bigg[c_e^2 \frac{m_e^2}{ s}\frac{1+\tau_X^2}{(1-\tau_X)(1-\cos^2\theta)}\\
&+c_e c_{\gamma} \frac{m_e^2}{2 s}\frac{(1-\tau_X)^2}{(1-\cos^2\theta)}+\frac{c_\gamma^2}{32}(1+\cos^2\theta)(1-\tau_X)^3\bigg] \notag \,,
\end{align}
up to corrections of order $m_e^2/s$. Here the first term is the contribution from the diagram with the axion coupling to the electron, the second is the interference term, and the last term is the contribution from the diagram with the axion radiated from the photon (axion-strahlung).
Appendix \ref{app:Amp_dets} describe the amplitude calculations in detail.
The differential cross-sections are shown in the left panel of Figure~\ref{fig:diffcs}. Both the hidden photon and the axion coupled to electrons are produced mostly for $\cos(\theta)\to 1$.
The angular distribution can therefore distinguish a new particle produced in t/u-channel diagrams (axions coupling to electrons or hidden photons) from particles produced in the s-channel (axions coupling to photons), though it isn't enough to distinguish within these categories.

We show the dependence of the differential cross-sections on the mass of the invisible state $m_X$ in the right panel of Figure~\ref{fig:diffcs}. The production via s-channel is suppressed for large $\tau_X$, but the production via t/u-channel is enhanced.
Hence, independently of the background present, the expected signal from the s-channel contribution reduces for larger masses, decreasing the experimental sensitivity to the axion coupling to photons.
We note that the t/u-channel contributions seem to diverge for both large angles and $\tau_X\to 1$, but are regularised by the electron mass.  

The chiral Belle programme is the proposal to use a polarized electron beam to collide with an unpolarized positron beam~\cite{Forti:2022mti}. Since the amplitude for hidden photon and axion-strahlung production is dominated by electrons and positrons with opposite helicities, whereas the production of axions through coupling to electrons is dominated by electrons and positrons with equal helicities, this won't have a qualitative effect on the differential production cross-sections. If \emph{both} the electron and positron beams are polarized, one can distinguish the hidden photon from an axion. In the case of a hidden photon, if electron and positron are polarized with \emph{equal} helicity, the signal would be suppressed by $m_e^2/s$ with respect to the unpolarized case. In contrast, the leading term for the cross-section for axions produced via electron couplings remains unchanged for electron and positron beams with equal helicities. 
These two production channels have the same angular distribution, but with the polarization of both beams one can significantly reduce the contribution of either channel, distinguishing them from one another. Similarly, in the context of a dark vector boson $Z_d$, which differs from the hidden photon considered by the inclusion of an axial-vector coupling, the angular distribution of the polarized differential cross-section of $e^+ e^- \to Z_d \gamma$ can be used to distinguish between the vector and axial-vector couplings~\cite{Lee:2020tpn}. 
We focus on longitudinal beam polarizations here, but for electron and positron beams with transversal polarization the information from the azimuthal angular distribution can be used to further discriminate between background and signal~\cite{Renard:1981es,Hikasa:1985qi,Ananthanarayan:2004xf}.

\begin{figure}[tbp]
\begin{center}
   \hspace{.5cm}  \begin{subfigure}[b]{0.2\textwidth}
         \centering
         \includegraphics[width=0.85\textwidth]{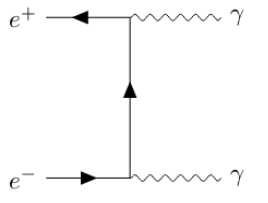}
     \end{subfigure}
     \hfill
     \begin{subfigure}[b]{0.2\textwidth}
         \centering
         \includegraphics[width=0.85\textwidth]{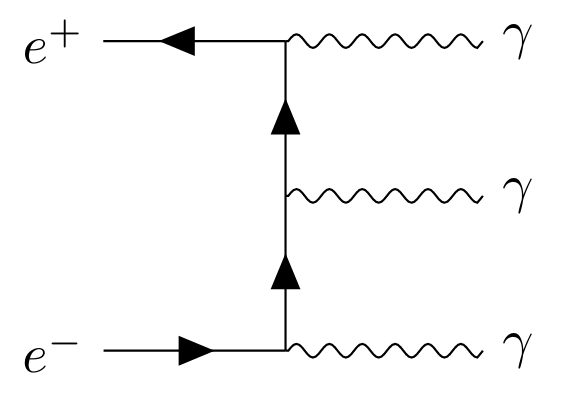}
     \end{subfigure}\\
     \hfill
    \begin{subfigure}[b]{0.2\textwidth}
         \centering
         \includegraphics[width=.9\textwidth]{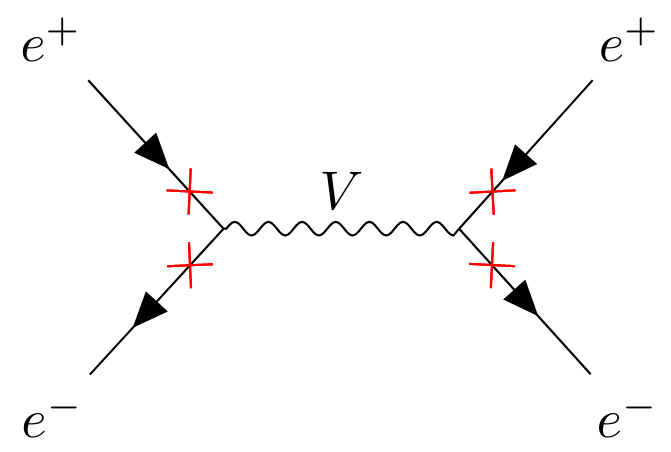}
     \end{subfigure}
     \hfill
     \begin{subfigure}[b]{0.18\textwidth}
         \centering
         \includegraphics[width=0.9\textwidth]{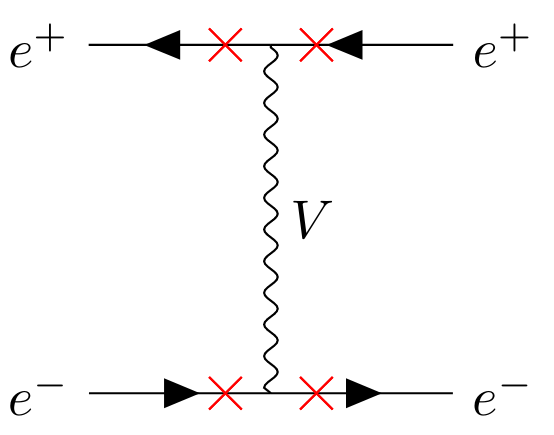}
     \end{subfigure}\\
     \hfill
     \begin{subfigure}[b]{0.18\textwidth}
         \centering
         \includegraphics[width=\textwidth]{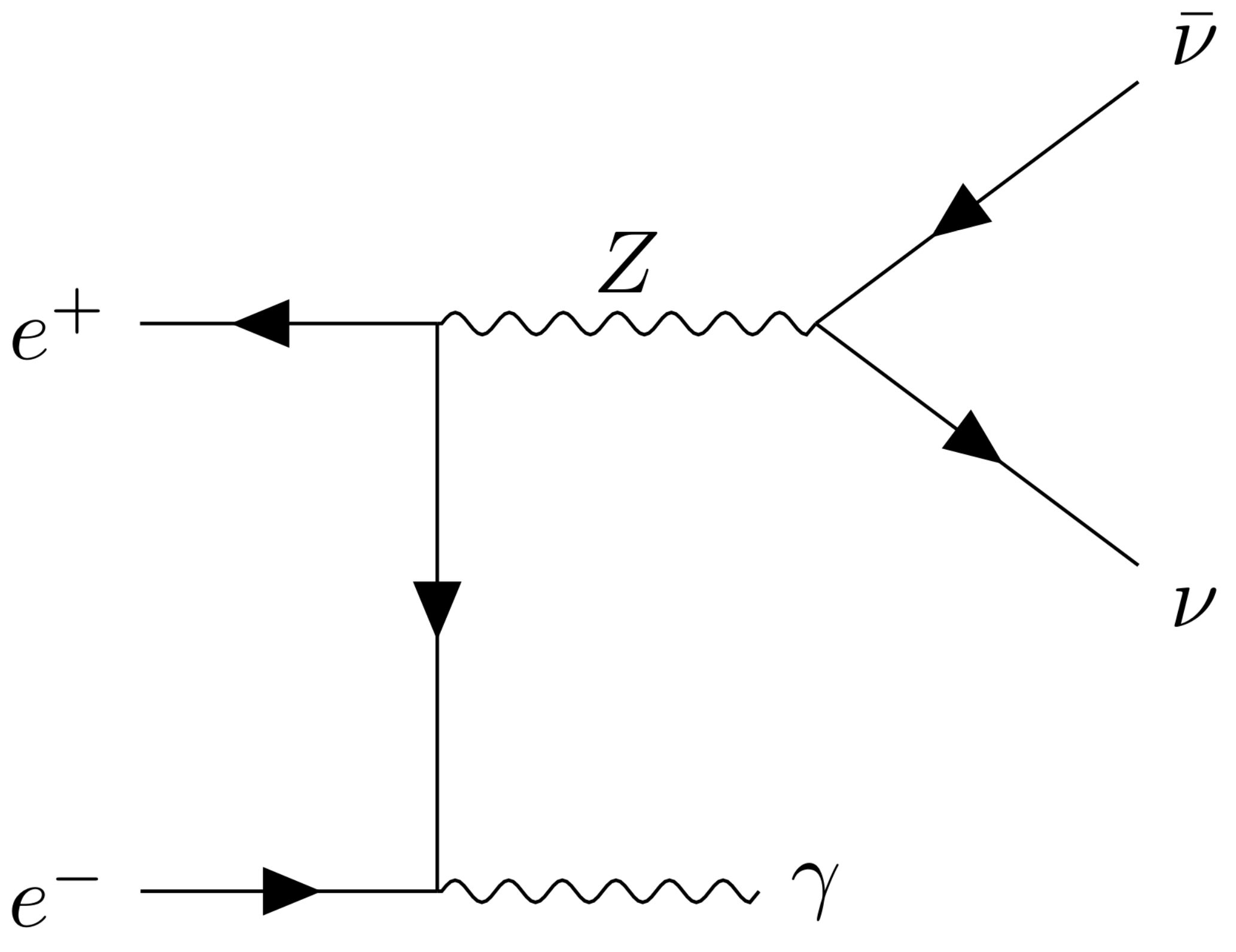}
     \end{subfigure}
     \hfill
     \begin{subfigure}[b]{0.18\textwidth}
         \centering
         \includegraphics[width=0.9\textwidth]{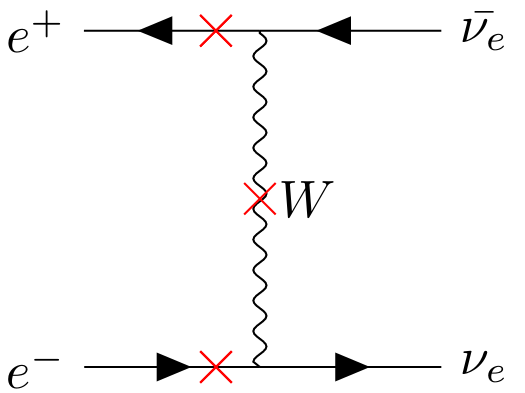}
     \end{subfigure}
     \end{center}
        \caption{Feynman diagrams for all background processes, $e^+ e^- \to \gamma \gamma (\gamma)$ (top), $e^+ e^- \to e^+ e^- \gamma (\gamma)$ (middle), and $e^+ e^- \to \nu \Bar{\nu} (\gamma)$ (bottom), with $V \in [\gamma, Z]$.}
        \label{fig:SM_back}
\end{figure}

%%%%%%%%%%%%%%%%%%%%%%%%%%%%%%%%%
%%%%%%%%%%%%%%%%%%%%%%%%%%%%%%%%%
\section{Event Generation} \label{sec:sims}
%%%%%%%%%%%%%%%%%%%%%%%%%%%%%%%%%
%%%%%%%%%%%%%%%%%%%%%%%%%%%%%%%%%

The signature in the production of an invisible state together with a photon is a single photon recoiling against missing energy. Any search for New Physics with this signature needs to account for a number of SM processes that produce final states that are difficult to distinguish. The SM background is dominated by the process $e^+ \, e^- \to e^+ \, e^- \, \gamma$ in which both the electron and the positron escape the detector. Additional background processes are $e^+ \, e^- \to \gamma \, \gamma \, (\gamma)$ in which one or two photons are lost, $e^+ \, e^- \to e^+ \, e^- \, \gamma (\gamma)$ in which the electron-positron pair isn't detected, and the irreducible production of neutrinos $e^+ \, e^- \to \nu \, \Bar{\nu} \, \gamma (\gamma)$. The selection of signal events takes advantage of the signal and background event distribution, following the analysis in~\cite{Belle-II:2018jsg} with modifications. 

 \subsection{Standard Model Background} \label{sec:Amp_SM} 

\begin{table}[b]
\begin{center}
\begin{tabular}{cp{1cm}|c}
\hline
SM process & &fraction\\[2pt]
\hline \hline
$e^+ e^-\gamma$ && 79.38\%\\[3pt]
$e^+ e^-\gamma\gamma$ &&10.39\%\\[2pt]
$\gamma\gamma\gamma$ && 9.72\%\\[2pt]
$\gamma\gamma$ && 0.51\%\\[2pt]
$\bar\nu \nu \gamma (\gamma)$ && $<0.01$\%\\[2pt]
\hline
\end{tabular}
\end{center}
    \caption{Fraction of the different SM backgrounds for simulations performed in this study.}
    \label{tab:SM_ratios}
\end{table}

\begin{figure}[bp]
    \centering
   \includegraphics[width=.49\textwidth]{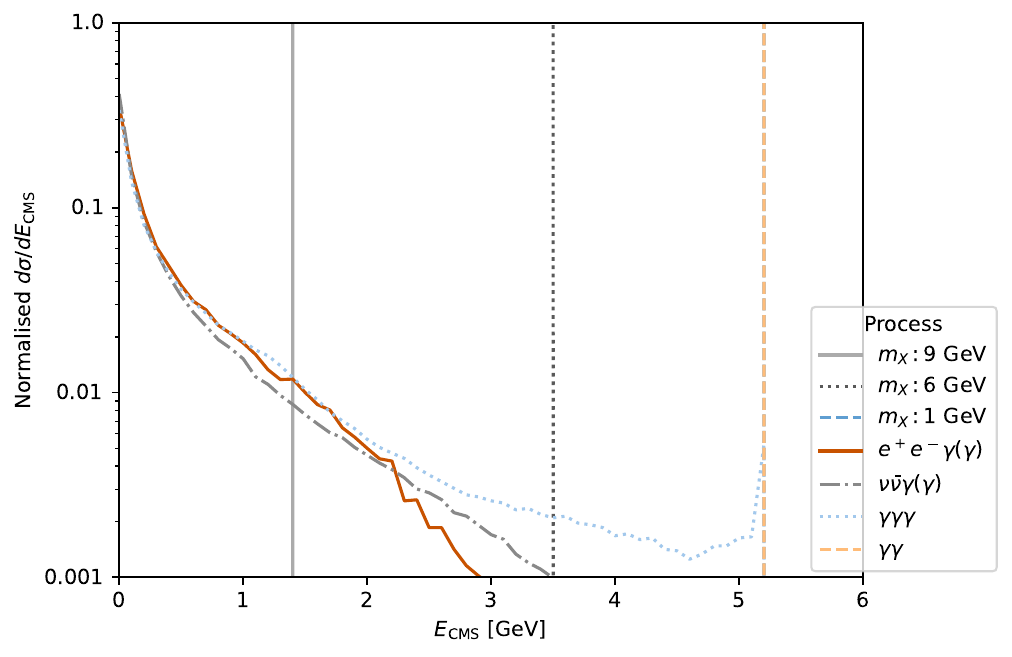}
    \caption{Comparison of the differential cross-section for the different SM background processes and the production of a new state with $m_X=1,6,9$ GeV as a function of the lab frame energy of the final state photon $E_\gamma$.}
     \label{fig:piechart}
  \end{figure}

Each SM background process has several contributions, where the corresponding Feynman diagrams are shown in Figure~\ref{fig:SM_back}.
Implementing minimal cuts on the lab frame energy $E_\gamma > 0.01$ GeV and asymmetric angular coverage $12.4^\circ \le \theta_\text{lab} \le 155.1^\circ$~\cite{USBelleIIGroup:2022qro}, we find that the fraction of the different processes contributing to the background shown in Table~\ref{tab:SM_ratios}.
In Figure~\ref{fig:piechart}, we show the normalised differential photon energy distribution of the different background processes compared to the signal process for three different masses of the invisible state $m_X=1,6, 9~$GeV using the cuts described above.
The shape of the differential energy distribution is independent, up to small corrections, of the spin or specific production process of the invisible state and shows how cuts on the energy can isolate the signal in particular for small masses $m_X$.
For high photon energies or small mass $m_X$, the background process $e^+e^- \to \gamma\gamma$ is increasingly important, whereas for low photon energies or high mass $m_X$ other background processes are more important. 
For this reason the BaBar search for hidden photons distinguishes two different mass regions, $-4 <m_X^2 < 36 \,\text{GeV}^2$ and $24 < m_X^2 < 69 $\, GeV$^2$\footnote{The upper limit is $m_X^2=63.5$ GeV$^2$ for the $\Upsilon (2S)$ dataset.} , in which the background from $e^+e^- \to \gamma\gamma$ and $e^+ \, e^- \to e^+ \, e^- \ \gamma $ dominate, respectively~\cite{BaBar:2008aby}. Similarly, Belle II anticipates two different signal regions for $m_X<6$ GeV and $m_X=6-8$ GeV~\cite{Belle-II:2018jsg}. In the following, we will analyse the proposed cuts for these analyses which are optimized to search for hidden photons and comment on the different optimization for axion searches and how polarized beams 
or a final state photon polarization measurement could improve the analysis.

  \begin{figure*}[tbp]
    \begin{center}
         \hspace{-.5cm}\includegraphics[width=1.\textwidth]{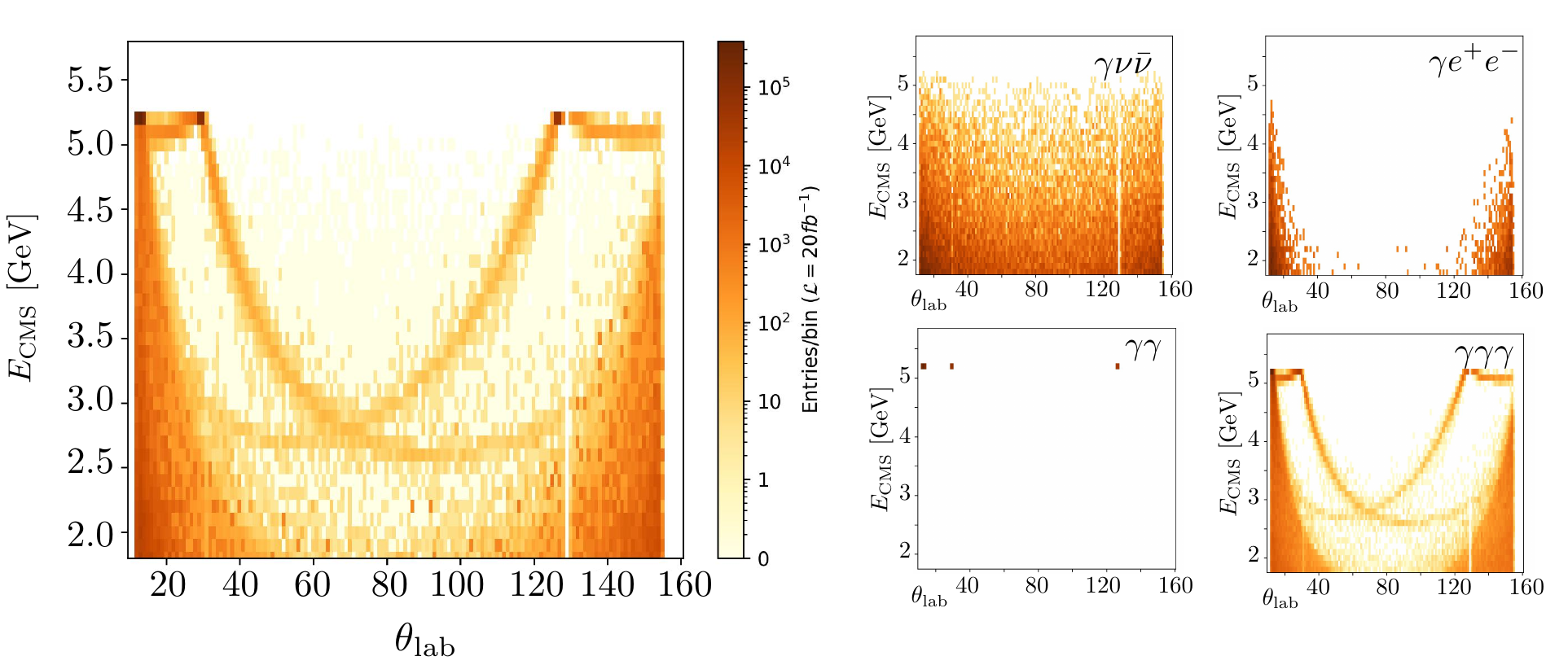}
     \end{center}
        \caption{Distribution of Standard Model background events for searches for final states with a photon and missing energy (left) and the separate contributions (right) displaying the shape of each contribution determined by the angular distribution and detector geometry, each with their own scaling (see text for further explanation).}
        \label{fig:SM_background}
\end{figure*}

% %%%%%%%%%%%%%%%%%%%%%%%%%%%%%%%%%
% %%%%%%%%%%%%%%%%%%%%%%%%%%%%%%%%%
% \section{Simulations} \label{sec:sims}
% %%%%%%%%%%%%%%%%%%%%%%%%%%%%%%%%%
% %%%%%%%%%%%%%%%%%%%%%%%%%%%%%%%%%
We simulate $5 \times 10^6$ background events using MadGraph \cite{Alwall:2014hca} with a minimum transverse momentum cut applied to outgoing photons (which is also applied to the invisible states),  $p_T \geq 0.01~\text{GeV}$ in order to avoid divergences. 
The simulations are performed for tree-level, fixed-order QED contributions with a simplified detector setup.
We, for example, do not consider the imperfections in the crystals in the detector, and the photon conversion probability.
The beam polarization is specified within MadGraph for $0-100\%$ polarized beams.
We require only one photon within the angular acceptance of the detector; either in the end-caps or the main barrel.
All other particles must be undetected, and specifically for all charged particles we also require for outgoing fermions to have $p_T \geq 0.1~$GeV. 

The distribution of photons from the various SM backgrounds in the plane spanned by the scattering angle $\theta_\text{lab}$ and centre of mass energy $E_\text{CMS}$ is shown in Figure~\ref{fig:SM_background}. The left panel shows the total background, and the right panels show the distributions for the different background processes. Note that the color coding applies only to the left panel. In the right panels, the color coding merely indicates the distribution of events within each panel, but the scaling varies for each panel. 

The panel with $\gamma \gamma$ has the same color coding as the total background on the left, whereas $\gamma \gamma \gamma$ and $e^+ e^- \gamma (\gamma)$ backgrounds are rescaled by a factor $~15-50$, and $\nu \Bar{\nu} \gamma (\gamma)$ by a factor of $5 \times 10^7$ with respect to the left panel in order to increase the visibility of the features.
For larger displays and individual color coding, see Appendix~\ref{app:SM_backs}.
The background from $\gamma\gamma$ final states has a fixed energy, but only contributes for certain angles because of two gaps between the endcaps and barrel detectors in the forward and backward direction, and the asymmetric angular coverage of the detector where only one photon is lost along the beampipe. As a consequence of the asymmetric beam energies and asymmetric angular coverage, processes in which photons are lost along the beampipe in the other direction don't contribute to the background because the recoiling photon isn't covered by the detector. 

The endcap gaps are also visible in the background from $\gamma\gamma\gamma$ final states, where the two bands crossing the central region of the plane correspond to photons lost in either forward or backward direction. The background from  $e^+e^-\gamma(\gamma)$ final states is most pronounced for small angles with the beampipe and there are fewer events for angles $40^\circ \lesssim \theta_\text{lab} \lesssim 120^\circ$, whereas background events from $\nu\bar \nu\gamma (\gamma)$ final states have an almost flat $\theta_\text{lab}$ distribution, but the contribution decreases for high photon energies. Our background distribution agrees with the unpublished results found in~\cite{Wakai:3598} where a detailed analysis of the different background processes is performed. We find more background events in the for large $\theta_\text{lab}$ Belle physics paper~\cite{Belle-II:2018jsg}, which includes higher-order QED corrections and performs a full detector analysis, though other cuts are applied. \\

\begin{figure*}[tbp]
    \centering
   \includegraphics[width=.5\textwidth]{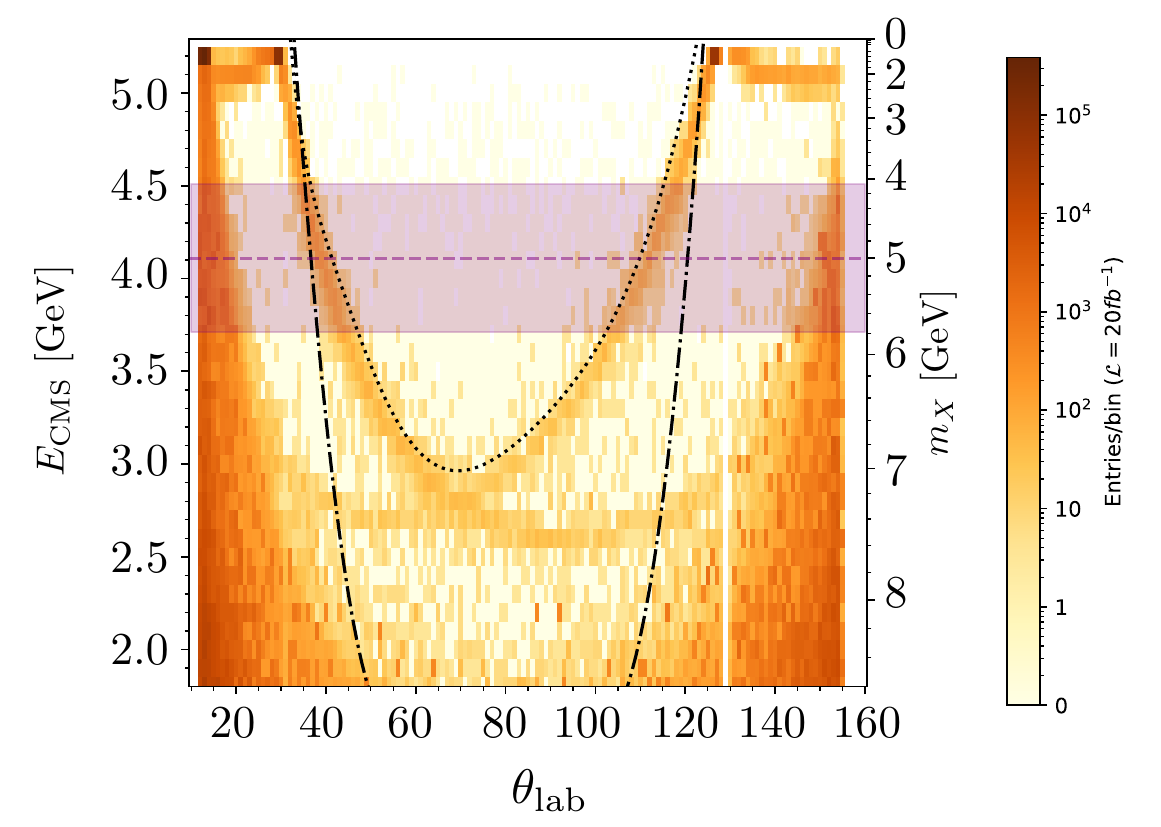}\includegraphics[width=.48\textwidth]{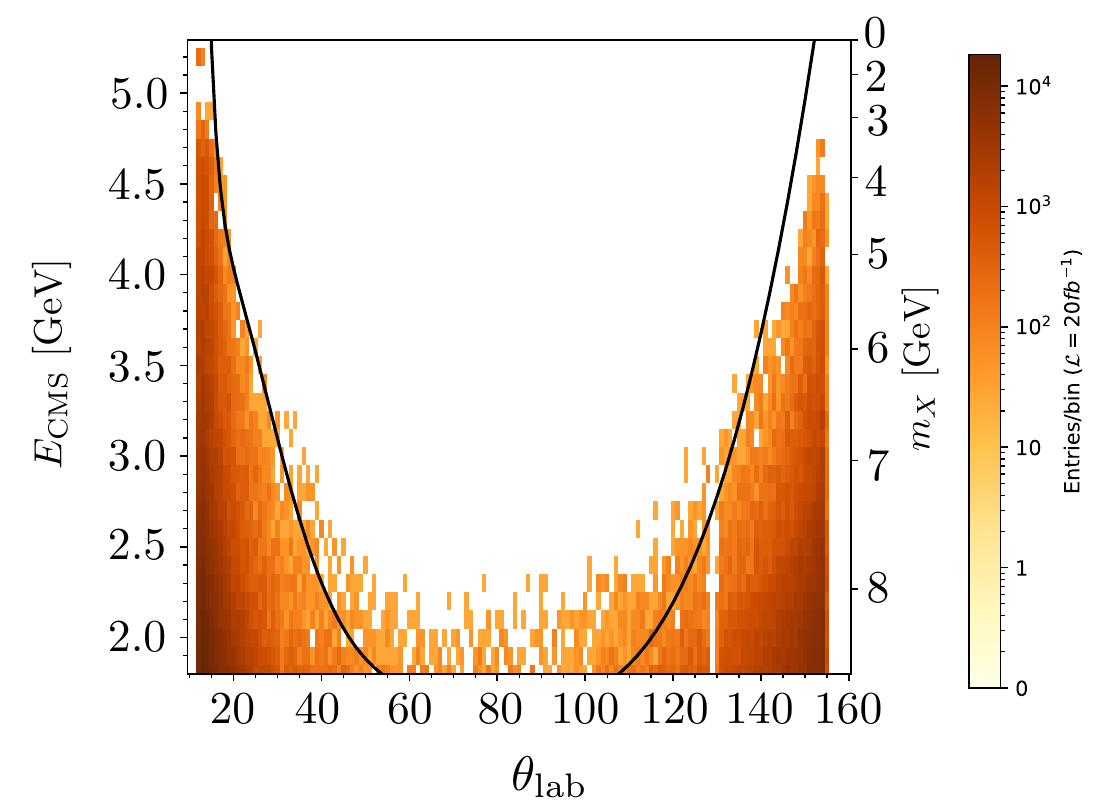}
    \caption{Event selection for unpolarized beams (left) and equal helicities polarized beams (right). In the left panel we also show the $E_\text{CMS}$ envelope used to define cuts in the search for a 5 GeV invisible state.}
     \label{fig:cuts}
  \end{figure*}

For electron and positron beams polarized with equal helicities, the backgrounds are significantly reduced. Backgrounds from $\gamma\gamma$ and $\gamma\gamma\gamma$ final states are substantially suppressed with respect to the unpolarized case, and the remaining background is mostly peaked towards small angles with the beam axis. In the right panel of Figure~\ref{fig:cuts}, we show the distribution of background events for beams polarized with the same helicities. Instead, for electron and positron beams polarized with opposite helicities the backgrounds are very similar to the unpolarized case. 
Hence in the following analysis, the case for oppositely polarized beams will not be considered. 

 \subsection{Signal} \label{sec:signal}
We implement the signal processes into MadGraph using UFO models \cite{Degrande:2011ua} based on modified Feynrules models for Axion-Like-Particles (ALPs)~\cite{Brivio:2017ije} and $Z'$ models~\cite{Basso:2008iv,Amrith:2018yfb,Deppisch:2018eth}.
The energy of the photon recoiling against the invisible state is fixed by the mass of the invisible state. The signal is therefore constant in $E_\text{CMS}$, and larger masses $m_X$ correspond to lower photon energies. The angular distribution of the signal peaks towards small and large angles with respect to the beam-line in the case of hidden photons or axions radiated off an electron, whereas it is approximately flat for axions coupled to photons. In case a signal is observed, and enough statistics are available, the angular distribution can be used to distinguish these models. 
  
We simulate $10^6$ events with the axion couplings fixed at $c_\gamma = 1$ and $c_{e} = 1 \times 10^4$ while the decay constant $f$ is kept as a free parameter. We assume that all hidden photons and axions leave the detector before they decay, or that they decay into invisible particles, so that their width is considered to be zero for the remainder of this analysis. This would be a good assumption if they represent mediators that dominantly decay into dark matter. We vary the parameters within,
\begin{gather}
    c_X \in [5\times10^{-6}, 1\times10^{-5},5\times10^{-5}, ..., 5\times10^{-3}]\,, \nonumber \\[5pt]
    f \in [4\times10^{5},2\times10^{5},4\times10^{4},...,4\times10^{2}]\,. \nonumber 
\end{gather}
The mass of the invisible state is varied with changing step sizes in order to increase precision; 
$m_X \in [1.0,8.0]$ GeV in steps of $1$ GeV with additional $m_X = 0.1,0.5, 8.5,$ and $9.0$ GeV.

For polarized beams, the signal changes as discussed in Section~\ref{sec:amp_form}. If the beams have the same helicities the background is substantially suppressed. The signal, in the case of a hidden photon, is chirally suppressed and very small compared to the unpolarized case. Similarly, axion coupled only to photons are produced with a strongly suppressed cross-section for beams with equal helicities. In contrast, the cross-section for the production of axions interacting with electrons isn't suppressed and one could take advantage of the lower background for polarized beams.

 \subsection{Event Selection} \label{sec:eventselection}
We distinguish searches for light new states $m_X<6$ GeV and heavy new states $m_X\geq 6$ GeV, because there is more background for softer photons, and the photon trigger efficiency varies significantly with energy~\cite{Belle-II:2018jsg}. We use the Belle II angular acceptance regions as described in ~\cite{Belle-II:2018jsg} and apply a cut on the energy of the detected photon, $E_{CMS} \geq 1.8$ GeV, which restricts the mass of the invisible state.
The angular coverage consists of three regions; the forward endcap $12.4^\circ < \theta < 31.4^\circ$, the mail barrel $32.2^\circ < \theta < 128.7^\circ$, and the backwards endcap $130.7^\circ < \theta < 155.1^\circ$.
  
In a first step, we impose an energy-dependent cut in the $\theta_\text{lab}-E_\text{CMS}$ plane, taking advantage of the correlation between energy and scattering angle for the distribution of the background events. 
The cut functions are polynomials that were fitted using an algorithm designed to minimise background events.
In contrast to the cut functions in \cite{Duerr:2019dmv}, our cuts reduce the background from  $\gamma \gamma \gamma$ and $\gamma \gamma$ final states. The details of the fit are explained in Appendix~\ref{app:ex_lims}.
In the left panel of Figure~\ref{fig:cuts}, the dotted black contour defines the cut for light invisible states, and the dash-dotted black contour defines the cut for heavy invisible states. The parameter space enclosed by these contours is the fiducial region. In the case of polarized beams with the same helicity, the energy-dependent $\theta_\text{lab}$-cut is shown by the black contour in the right panel of Figure~\ref{fig:cuts}. The reduced background allows for a large fiducial region compared to the unpolarized case, so that the separation for different mass regions is unnecessary.\\ 
In a second step, we introduce a mass-dependent cut on $E_\text{CMS}$. The fixed relation between the final state photon energy and the mass of the invisible state is shown by the alternative y-axis in Figure~\ref{fig:cuts}. It allows a more targeted search since signal events are predicted by the invisible state mass. 
The inclusion of higher-order corrections causes a smearing in the final state photon energy, and hence in addition to the energy-dependent $\theta_\text{lab}$-cut we only select events in a window of $m_X\pm 0.4 $ GeV. This window is shown in the left panel of Figure~\ref{fig:cuts} for the case of $m_X=5$ GeV, which is wider than the one derived in the full Belle analysis based on the Novosibirsk function for higher masses $m_X$~\cite{Belle-II:2018jsg} and comparable to the energy window used in~\cite{Duerr:2019dmv}. We checked that it captures more than $95\%$ of the signal for the example given in~\cite{Belle-II:2018jsg} for $m_X=7 $ GeV.\\

\begin{figure}[bp]
    \centering
    \includegraphics[width=0.49\textwidth]{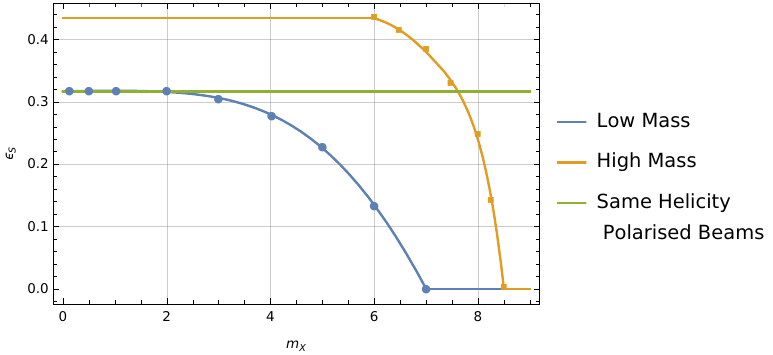}
    \caption{Trigger efficiency for low- and high mass region and the fixed efficiency used for the analysis with polarized beams.}
    \label{fig:efficiency}
\end{figure}

The trigger efficiency $\epsilon_s$ is taken from~\cite{Belle-II:2018jsg} and interpolated for different masses as shown in Figure~\ref{fig:efficiency}. In the case of polarized beams, we make the conservative choice to use the trigger efficiency for the low-mass region, which is worse than the efficiency expected for the high-mass region as long as $m_X< 8$ GeV. 

\begin{figure*}[tbp]
    \centering
    \begin{subfigure}[b]{0.49\textwidth}
        \centering
        \includegraphics[width=\textwidth]{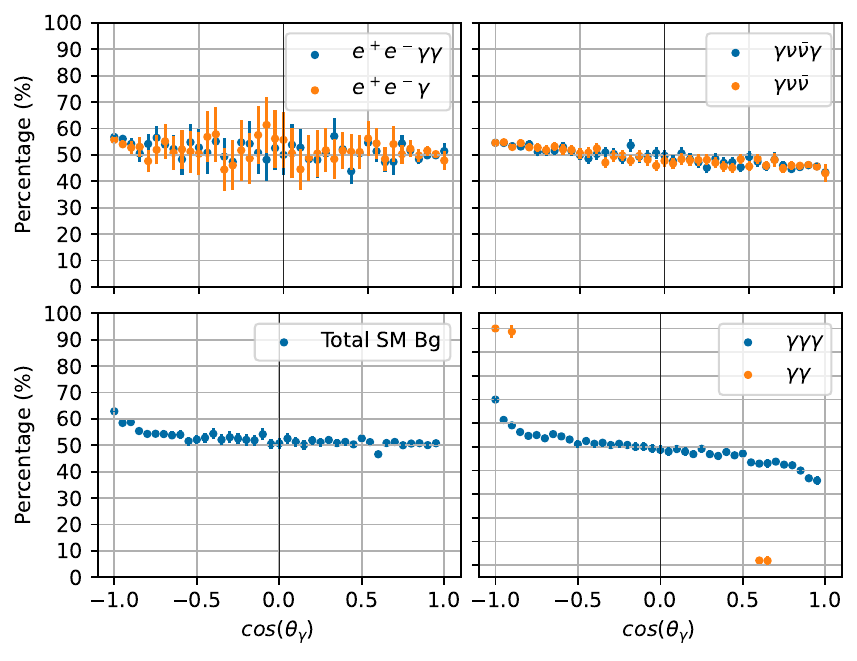} 
    \end{subfigure}
    \begin{subfigure}[b]{0.49\textwidth}
        \centering
        \includegraphics[width=\textwidth]{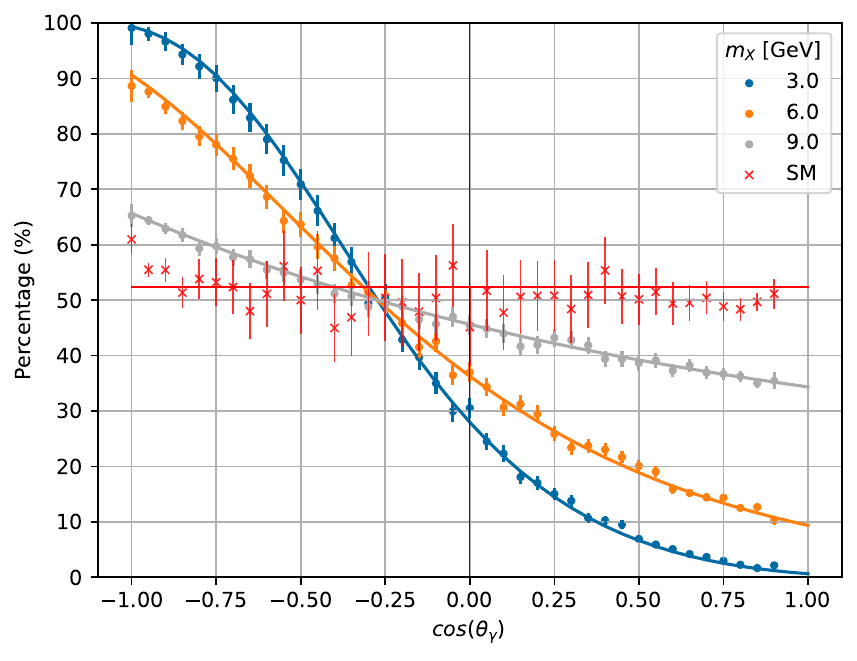}
    \end{subfigure}

    \caption{Helicity fraction distribution of $2\lambda_{e^-} = \lambda_\gamma$ for background processes (left) and hidden photon (right) for $m_X = 3.0, 6.0,$ and $9.0$ GeV as a function of the polar angle $\cos{\theta_\gamma}$.}
    \label{fig:hel_cos}
\end{figure*}

%%%%%%%%%%%%%%%%%%%%%%%%%%%%%%%%%
%%%%%%%%%%%%%%%%%%%%%%%%%%%%%%%%%
\section{The Polarization of the Final state Photon} \label{sec:photonpolarization}
%%%%%%%%%%%%%%%%%%%%%%%%%%%%%%%%%
%%%%%%%%%%%%%%%%%%%%%%%%%%%%%%%%%

We discussed how the angular distribution together with the polarization of the incoming beams can be used to distinguish between the hidden photon, axion coupling to photons, and axion coupling to electron contributions.
The polarization of the beams can be used to significantly reduce the background for the axion coupling to electrons, but leaves the background for the other two dark matter contributions essentially unchanged.
We therefore will now consider how the helicity of the outgoing photon can be used as a complementary discriminator.

Figure \ref{fig:hel_cos} displays the percentage of outgoing photons with helicity equivalent to that of the incoming electron for $100\%$ polarized beams, a detailed description of the results can be found in Appendix \ref{app:hel_dia}.
The error-bars shown are calculated using 
\begin{equation}
    D_i = \frac{N_i}{N_{\text{tot}}} \hspace{+0.5cm} \Rightarrow \delta D_i = \frac{N_i}{N_{\text{tot}}} \sqrt{\frac{1}{N_i} + \frac{1}{N_{\text{tot}}}}\,,
\end{equation}
for $N_i$ number of events for each $\cos{\theta}$ and total number of events $N_{\text{tot}}$.
The distribution of photon helicity for the standard model background is centered around $50 \%$ with limited angular dependence as seen on the bottom left panel. 
Any significant variation from this would indicate the presence of New Physics.

The hidden photon production has the photon helicity matching the helicity of the fermion travelling in the same direction, which is featured in the s-shape distribution seen on the panel on the right.
For small $m_X$, this correlation is more pronounced, but as $m_X$ increases, the direction of the photon becomes irregular and its helicity more independent.
Therefore, the overall fraction tends towards $50 \%$ for increasing $m_X$.

The helicity distribution for axion coupling to photons has the same s-shape, though the mass of the axion has no influence on the result as it factorises out and becomes a part of the coupling constant for the process. 
The s-shape helicity distribution is substantially different to the background and can therefore be used to distinguish the signal and provide improved exclusion limits.
We will not carry this out due to the unknown detector setup needed to measure the helicity of the outgoing photon.

As the electron and positron helicities are the same for the axion coupling to electrons, the photon helicity will match the helicity of both or neither, and the resulting helicity fraction has no angular dependence.
For small $m_a$, the photon helicity will match the incoming fermions, resulting in a constant percentage of $100 \%$. 
But like the hidden photon process, as the axion mass increases, the distribution approaches $50 \%$ with increasing $m_a$.
The axion coupling to electrons will cause a shift, and hence it can easily be distinguished from the other to dark matter contributions, but would be difficult to separate from the background.

\begin{figure*}[tbp]
    \centering
    \includegraphics[width=1\textwidth]{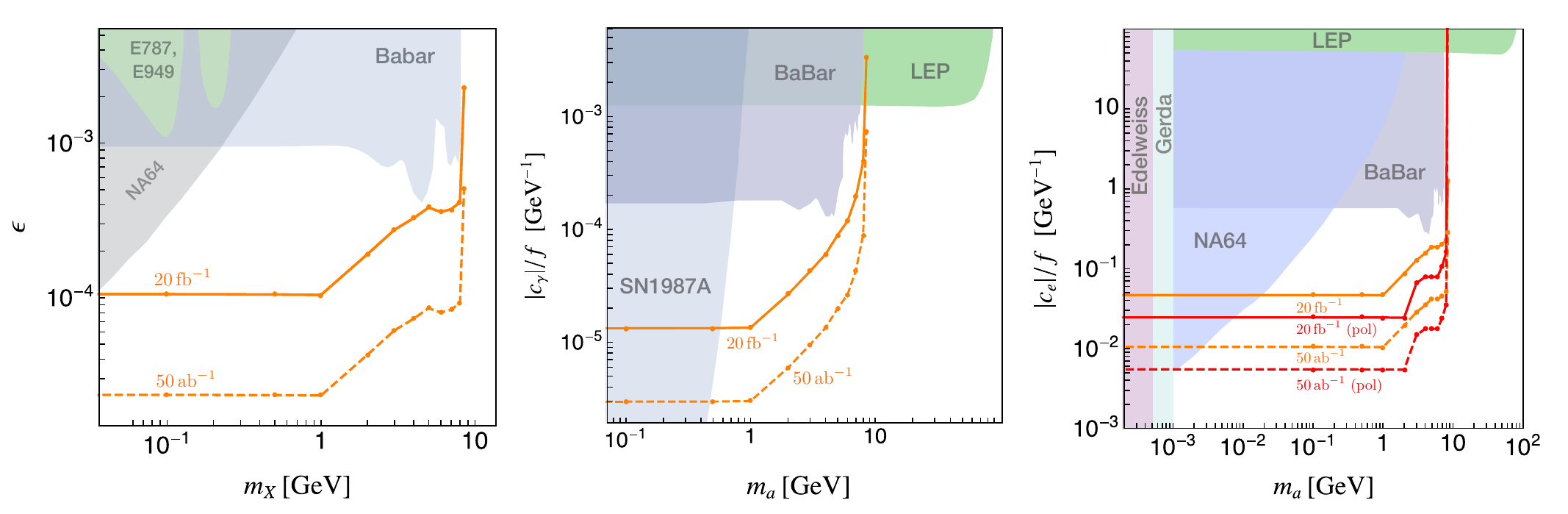}
    \caption{Belle II sensitivity to hidden photons (left), axions coupling to photons (center) and axions coupling to electrons (right). Details are in the text.}
    \label{fig:NPlimits}
\end{figure*}

\subsection*{Photon polarization}
In addition to the beam polarisation the polarisation of the final state photon can be used to determine the dirac structure in the production amplitude. This has been successfully used to examine of chiral structure of the operator responsible for $b \to s \gamma$ transitionsat LHCb~ \cite{LHCb:2021byf}. Measuring the final state photon polarization is extremely challenging, but if it's possible it could provide an additional handle on the spin of the dark matter state. We define
\begin{align}
\alpha_\gamma =\frac{\gamma_L-\gamma_R}{\gamma_L+\gamma_R}\,,
\end{align}
where $\gamma_{L/R}$ is the number of left- and right-handed photons respectively, and the photon polarization is therefore the ratio of the two polarization states.
For the SM background, $\alpha_\gamma = 0.5$ is expected for random photon polarization.
If only one beam is polarized, all dark matter contributions will have the same photon polarization as the background, but this result can be significantly changed when polarizing both beams.
When the incoming beams have opposite helicities, one can achieve $|\alpha_\gamma| \approx 0.8$  for axion-strahlung and hidden photon when implementing forwards/backwards angular cuts.
Whereas for the axion coupling to electrons with incoming beams with equal helicity, for small axion masses regardless of angular cuts one finds $| \alpha_\gamma | \approx +1$.
See appendix \ref{app:Amp_dets} for further detail.

%%%%%%%%%%%%%%%%%%%%%%%%%%%%%%%%%
%%%%%%%%%%%%%%%%%%%%%%%%%%%%%%%%%
\section{Sensitivity to New Physics} \label{sec:NP}
%%%%%%%%%%%%%%%%%%%%%%%%%%%%%%%%%
%%%%%%%%%%%%%%%%%%%%%%%%%%%%%%%%%

In the following we discuss the sensitivity reach of Belle II for hidden photons and axions interacting with photons or electrons for unpolarized and polarized beams. We assume that the new states are stable on collider scales, for example because they decay into dark matter. We obtain the expected 90\% CL upper limit on the observed number of signal events $\mu_S$ by demanding that the Poisson probability of observing not more than the number of background events $ \mu_B $ if $\mu_B+\mu_S$ events are expected is $P(\mu_B,\mu_S+\mu_B)>0.1$ as in~\cite{Belle-II:2018jsg}.\\

Figure~\ref{fig:NPlimits} shows the sensitivity reach of Belle II for 20 fb$^{-1}$ (orange contour) and 50 ab$^{-1}$ (dashed orange contour) for unpolarized beams and the improved sensitivity in the case of an axion coupling to electrons using polarized beams is shown by red contours.

For comparison, we show limits from a search for hidden photons by BaBar~\cite{BaBar:2017tiz}, limits from beam-dumps experiments E787, E949~\cite{Davoudiasl:2014kua,Essig:2013vha, E787:2004ovg, BNL-E949:2009dza} and 
NA64~\cite{Banerjee:2019pds}. 
The centre plots shows BaBar limits on axions from single-photon decays of $\Upsilon(1)$~\cite{BaBar:2008aby} and mono-photon searches by LEP~\cite{Fox:2011fx,DELPHI:2008uka}, taken from~\cite{Dolan:2017osp}, as well as constraints from 
the supernova SN1987A~\cite{Chang:2016ntp, Lucente:2021hbp}. 
In the right panel we show the bounds from the neutrinoless double-beta decay experiment Gerda~\cite{GERDA:2020emj}, the helioscope Edelweiss~\cite{EDELWEISS:2018tde} taken from~\cite{Bauer:2021mvw}, together with the bounds by NA64~\cite{NA64:2020qwq}, BaBar~\cite{BaBar:2017tiz} and LEP~\cite{DELPHI:2008uka}, taken from~\cite{Darme:2020sjf}. Note that the bounds from Edelweiss and Gerda require the axion to be stable on astrophysical scales and therefore only constrain $m_a< 2m_e$. \\
The projected sensitivity reach exceeds the existing constraints for all three models by about an order of magnitude at least. In all models the sensitivity drops with larger mass $m_X$ for which backgrounds are larger. For the case of axion-strahlung shown in the centre panel, this drop is particularly steep, because the cross-section drops for large masses as shown in the right panel of Figure~\ref{fig:diffcs}. The slight improvement of sensitivity for masses $m_X>6$ GeV is a consequence of the better trigger sensitivity. Note that the projected sensitivity for masses $m_X<2$ GeV at $50$ ab$^{-1}$ can only be achieved if backgrounds from cosmic rays are fully understood~\cite{Belle-II:2022cgf}.  \\
Our results can be compared with previous analyses of the sensitivity reach of Belle II. Our analysis largely follows~\cite{Belle-II:2018jsg}, that also sets constraints on an invisible spin 1 state with the main differences that we implement different $\theta_{\text{lab}}-E_\text{CMS}$ cut functions and a constant mass-dependent window as opposed to previous estimates in~\cite{Belle-II:2018jsg} that consider higher-order effects and a more sophisticated detector simulation.
As a consequence, we obtain projections roughly a factor three better. An analysis that translates this projection for the case of axions produced in axion-strahlung can be found in~\cite{Dolan:2017osp, Darme:2020sjf}. The improvement with the method we used is again roughly a factor three in the projected sensitivity, consistent with the hidden photon case. A recast for axions interacting with electrons has been performed in~\cite{Darme:2020sjf}, and we find again that our projections are roughly 3–4 times better using the improved cuts.  \\
If a signal is observed at Belle II, it is possible in principle to determine whether the invisible state is a vector boson ($c_X$), an axion interacting with photons ($c_\gamma$), or an axion interacting with electrons ($c_e$). The angular distribution of the signal events can distinguish between $c_\gamma$ and ${c_e,c_X}$, whereas the beam polarization suppresses the signal for ${c_X,c_\gamma}$ and doesn't affect the signal for ${c_e}$.

%%%%%%%%%%%%%%%%%%%%%%%%%%%%%%%%%
%%%%%%%%%%%%%%%%%%%%%%%%%%%%%%%%%
\section{Conclusions} \label{sec:conc}
%%%%%%%%%%%%%%%%%%%%%%%%%%%%%%%%%
%%%%%%%%%%%%%%%%%%%%%%%%%%%%%%%%%

We propose measurement strategies that can be used at electron positron colliders to determine the spin, mass, and production mechanism of an invisible state produced in association with a photon $e^+e^-\to \gamma + X$. In particular, hidden photons and axions that interact with electrons or photons can be distinguished even in the absence of detecting any of their decay products.\\
The angular distribution of the final state photon is sensitive to the production mechanism and can distinguish s-channel production as in the case of axion-strahlung from t/u-channel production, e.g. of a hidden photon or an axion produced from interactions with $e^+$ or $e^-$ directly. If both electron and positron beams can be polarized with equal helicities, the hidden photon cross-section is strongly suppressed with respect to the unpolarized cross-section, and only the cross-section for axions interacting with electrons remains unchanged. In combination, the angular distribution and the beam polarization can distinguish between these three models. Further, the dependence of the polarization of the final state photon on the polar angle can be used to discriminate between the different models as well as the SM background if it can be reconstructed with future detectors. For beams with opposite helicities and  small masses $m_X$, both the hidden photon and axion coupling to photons helicity fraction distributions are significantly different from the SM background. While beams with equal helicities can be used to reduce backgrounds for axions interacting with electrons, the sensitivity for both axions interacting with photons as well as hidden photons can be increased if final state photon polarization can be measured. \\

The SM background is also significantly reduced if both electron and positron beams are polarized. A careful analysis of different SM background processes shows that a combination of a universal cut in the plane spanned by the scattering angle $\theta_\text{lab}$ and centre of mass energy $E_\text{CMS}$ together with a mass-dependent cut could improve the projections for searches at Belle II, assuming the method described in this paper accounts for smearing effect from higher-order QED.
If a run with polarized beams can be performed the sensitivity to axions interacting with electrons is further improved by a factor two. We compare the projections with constraints from other experiments and astrophysics and identify the parameter space that can be probed at Belle II using the proposed measurement strategy.

%%%%%%%%%%%%%%%%%%%%%%%%%%%%%%%%%
%%%%%%%%%%%%%%%%%%%%%%%%%%%%%%%%%
\section{Acknowledgements} \label{sec:ack}
%%%%%%%%%%%%%%%%%%%%%%%%%%%%%%%%%
%%%%%%%%%%%%%%%%%%%%%%%%%%%%%%%%%
We thank Christopher Hearty for helpful comments on the distribution of background events and the geometry of the Belle II detector.

%%%%%%%%%%%%%%%%%%%%%%%%%%%%%%%%%
%%%%%%%%%%%%%%%%%%%%%%%%%%%%%%%%%
\appendix
%%%%%%%%%%%%%%%%%%%%%%%%%%%%%%%%%
%%%%%%%%%%%%%%%%%%%%%%%%%%%%%%%%%

\section{Amplitude Calculations Details} \label{app:Amp_dets}
In the following we will describe how the amplitudes and analytical results were derived, for further detail see \cite{Erner:2024}. 
The general helicity amplitude for the process, 
\begin{equation}\label{eq:proc_1}
    e^+(p_1,\lambda_{e^+}) + e^-(p_2,\lambda_{e^-}) \to \gamma(q_1,\lambda_\gamma) + X(q_2,\lambda_X)\,,
\end{equation}
with
\begin{align}\label{eq:amp_1}
    \mathcal{M} &= \mathcal{M}_\mu \epsilon^\mu(q_1,\lambda_\gamma)\notag\\ 
    &= \Bar{v}(p_{e^+},\lambda_{e^+})\Gamma_\mu u(p_{e^-},\lambda_{e^-})\epsilon^\mu(q_1,\lambda_\gamma)\,,
\end{align}
where $\Gamma_\mu$ is the current for the interaction.
Belle II has incoming electron and positron antiparallel along the z-axis with energies $E_1$ and $E_2$ respectively. 
Their momenta are defined, 
\begin{align}
    p_1^\mu &= \left(E_1,0,0,-\sqrt{E_1^2 - m_e} \right)\,, \\
    p_2^\mu &= \left(E_2,0,0,\sqrt{E_2^2 - m_e} \right)\,, \\
    q_1^\mu &= E_\gamma (1,\cos{\phi} \sin{\theta},\sin{\phi} \sin{\theta}, \cos{\theta})\,, \\
    q_2^\mu &= p_1^\mu + p_2^\mu - q_1^\mu\,,
\end{align}
together with the two outgoing particles (photon and invisible state $X$) where $E_\gamma$ is the photon energy in the lab frame, $\theta$ is the angle between the photon and incoming beams, and $\phi$ between the photon and the x-axis in the transverse plane.

\subsection*{Fermion Spin Vectors}
Spin-$1/2$ particles with four-momentum $p^\mu = (E, \boldsymbol{p})$ have corresponding spin four-vector where $\lambda$ is the helicity of the particle with $\lambda = \pm 1/2$ \cite{Haber:1994pe},
\begin{align}
    &\text{Massive}: \, S^\mu = \frac{2 \lambda}{m} \left( |\boldsymbol{p}|, E \hat{\boldsymbol{p}} \right) \label{eq:spin_vec_massive}\,, \\
    &\text{Massless}: \, S^\mu = 2 \lambda (1,\hat{\boldsymbol{p}} ) \,.\label{eq:spin_vec_massless}
\end{align}
Helicity projection expression can be written as,
\begin{align}
    u(p,\lambda) \Bar{u}(p,\lambda) &= \frac{1}{2} \, (1+\gamma^5\cancel{S})(\cancel{p} + m)\,, \\
    \underset{m \to 0}{\Rightarrow}  u(p,\lambda) \Bar{u}(p,\lambda) &= \frac{1}{2} \, (1+2\lambda_\gamma^5)\cancel{p}\,, \\
    v(p,\lambda) \Bar{v}(p,\lambda) &= \frac{1}{2} \, (1+\gamma^5\cancel{S})(\cancel{p} - m)\,, \\
    \underset{m \to 0}{\Rightarrow}  v(p,\lambda) \Bar{v}(p,\lambda) &= \frac{1}{2} \, (1-2\lambda_\gamma^5)\cancel{p}\,,
\end{align}
which are used to implement the fermion spin vector into the amplitude.
Define amplitudes for right-and left-handed fermions,
\begin{align}
    &|\mathcal{M}_{RR}|^2 = |\mathcal{M}|^2(\lambda_{e^-} = +1,\lambda_{e^+} = +1) \nonumber\,,\\
    &|\mathcal{M}_{LR}|^2 = |\mathcal{M}|^2(\lambda_{e^-} = -1,\lambda_{e^+} = +1) \nonumber\,, 
\end{align}
similarly for $|\mathcal{M}_{LL}|^2$ and $|\mathcal{M}_{RL}|^2$.
For longitudinally polarised fermion beams, the amplitude can be separated into parts proportional to the four combinations of fermion helicity,
\begin{align}
    |\mathcal{M}|^2 = \frac{1}{4} \biggl\{ &\left(1 + P_{e^-} \right)\left(1 + P_{e^+} \right) |\mathcal{M}_{RR}|^2 \label{eq:M_e_pol} \\
    & + \left(1 - P_{e^-} \right)\left(1 - P_{e^+} \right) |\mathcal{M}_{LL}|^2 \nonumber \\
    & + \left(1 + P_{e^-} \right)\left(1 - P_{e^+} \right) |\mathcal{M}_{RL}|^2 \nonumber \\
    & + \left(1 - P_{e^-} \right)\left(1 + P_{e^+} \right) |\mathcal{M}_{LR}|^2 \biggr\}\,, \nonumber
\end{align}
where $P_{e^\pm}$ is the degree of electron and positron polarization,
\begin{equation}\label{eq:f_pol}
    P_{e^{\pm}} = \frac{n_{e^{\pm}_R} - n_{e^{\pm}_L}}{n_{e^{\pm}_R} + n_{e^{\pm}_L}}\,,
\end{equation}
and $n_{e^\pm_{R/L}}$ denote the number of left- and right-handed electrons and positrons in each beam. An unpolarised beam has $P_{e^\pm}=0$, and $P_{e^\pm}=\pm 1$ are $100 \%$ left- and right-handed polarized beams respectively \cite{Moortgat-Pick:2005jsx}. 

\subsection*{Photon polarization Vectors}
A spin-$1$ particle with four-momentum $k^\mu = (k_0, \boldsymbol{k})$ and helicity $\lambda$, moving in arbitrary direction has polarization four-vector basis,

\begin{align}
    \epsilon^\mu_1 (k) &= \frac{1}{|\boldsymbol{k}|k_T} \left( 0, k_x k_z, k_y, k_z, - k_T^2 \right)\,, \\
    \epsilon^\mu_2 (k) &= \frac{1}{k_T} \left( 0, -k_y, k_x, 0 \right)\,, \\
    \epsilon^\mu_3 (k) &= \frac{k_0}{|\boldsymbol{k}|\sqrt{k^2}} \left( \frac{\boldsymbol{k}^2}{k_0}, k_x, k_y, k_z \right)\,, \label{eq:pol_long} \\
    \epsilon^\mu_4 (k) &= \frac{1}{\sqrt{k^2}} \left( k_0, k_x, k_y, k_z \right)\,,
\end{align}
where $k_T = \sqrt{k_x^2 + k_y^2}$ \cite{Hagiwara:1987}.

The longitudinal polarization vector for a boson with $\lambda=0$ is described by eq.~\eqref{eq:pol_long} and helicity eigenvectors with $\lambda=\pm 1$ are,
\begin{equation}
    \epsilon^\mu ( k, \lambda ) = \frac{1}{\sqrt{2}} \biggl[ - \lambda \epsilon^\mu_1 (k) - i \epsilon^\mu_2 (k) \biggr]\,. \label{eq:pol_vec}
\end{equation}

The amplitude can be separated using the fraction of left- to right-handed outgoing photons $P_\gamma$,

\begin{align}
    |\mathcal{M}|^2 &= \frac{1}{2} \left\{ \left(1 + P_\gamma \right)\gamma_R +\left(1 - P_\gamma \right) \gamma_L \right\}\,, \label{eq:M_p_pol}
\end{align}

 where $\gamma_L = |\mathcal{M}|^2(\lambda_\gamma=-1)$ and $\gamma_R = |\mathcal{M}|^2(\lambda_\gamma=+1)$.

\subsection{Hidden Photon}
The amplitude of the hidden photon production has two contributions,
\begin{align}
    \mathcal{M}_1 = &\frac{e \, c_X}{t - m_e^2}\epsilon_\mu (q_1,\lambda_\gamma) \epsilon_\nu (q_2,\lambda_X) \\
    & \hspace{5ex} \Bar{v}(p_2,\lambda_{e^+}) \gamma^\mu (\cancel{p_1} - \cancel{q_1} + m_e) \gamma^\nu u(p_1,\lambda_{e^-})\,, \nonumber \\
    \mathcal{M}_2 = &\frac{e \, c_X}{u - m_e^2} \epsilon_\nu(q_1,\lambda_\gamma) \epsilon_\mu (q_2,\lambda_X) \\
    & \hspace{5ex} \Bar{v}(p_2,\lambda_{e^+}) \gamma^\mu (\cancel{p_1} - \cancel{q_2} + m_e) \gamma^\nu u(p_1,\lambda_{e^-})\,. \nonumber
\end{align}
Using FeynCalc \cite{Shtabovenko:2020gxv}, the matrix amplitude squared is calculated using the momenta, spin and polarization vectors described previously.
As the hidden photon is undetectable, we will sum over its spin using its mass $m_X$.
The leading term $\mathcal{O}(m_e^0)$ of the final total amplitude is given below in eq.~\eqref{eq:DP_amp},
where,
\begin{gather}
    E_\pm = E_1 \pm E_2\, \hspace{+1cm} \text{and} \hspace{+1cm} \beta_X^2=1 + \tau_X^2 \,.
\end{gather}
Two separate expressions containing the incoming fermion helicities emerge; $(\lambda_{e^-}-\lambda_{e^+})$ and $(\lambda_{e^-}\lambda_{e^+}-1)$ which both go to zero for $\lambda_{e^+} = \lambda_{e^-}$. Hence the helicities are required to be opposite and any dependence on them being equal are of order $\mathcal{O}(m_e^2)$.
The effects of the polarization of the incoming beams on the amplitude is investigated using,
\begin{equation}
    \mathcal{R}\left(P_{e^-},P_{e^+}\right) = \frac{|\mathcal{M}|^2 \left( P_\gamma = 0 \right)}{|\mathcal{M}|^2 \left( P_\gamma = P_{e^\pm} = 0 \right)}\,,
\end{equation}
where $\mathcal{R} \left(P_{e^-},P_{e^+}\right) = (1-P_{e^-} P_{e^+})$ at $\mathcal{O}(m_e^0)$ and the cross-section is maximally enhanced for fully oppositely polarised beams. 
The photon polarization describes the ratio of right-handed to left-handed photons,
\begin{equation}
    \alpha_\gamma = \frac{\gamma_L - \gamma_R}{\gamma_L + \gamma_R}\label{eq:phot_pol_2} \,,
\end{equation}
which at order $\mathcal{O}(m_e^0)$ is found to be,
\begin{align} \label{eq:a_g_DP}
    \alpha_\gamma=\frac{\left(P_{e^-}-P_{e^+}\right)}{\left(P_{e^-}P_{e^+}-1\right)}
    f\left(E_1,E_2,m_X\right)\,,
\end{align}
where largest $\alpha_\gamma$ is found for small $m_X$ and fully oppositely polarised beams.
As seen in the helicity fraction distributions for the hidden photon, see section \ref{sec:photonpolarization}, it is possible to significantly affect the photon polarization by applying angular cuts in the forwards or backwards directions with $|\alpha_\gamma| \ge 0.8$ for $m_X \le 4$ GeV.
\onecolumngrid
\begin{align}
    |\mathcal{M}|^2 &= -\frac{c_X^2 e^2 \csc^2{\theta} }{2 s (\tau_X-1)^2}
    \Biggl\{ \beta_X ^2 \lambda_\gamma  (\lambda_{e^-}-\lambda_{e^+}) \left[ E_{+} E_{-} (\cos (2 \theta )+3)+\left(4 E_{+}^2-2 s\right) \cos{\theta}\right] \label{eq:DP_amp} \\
    &\hspace{+5ex} -2 (\lambda_{e^-} \lambda_{e^+}-1) \left[ -2 \beta_X ^2 E_{+} E_{-} \cot{\theta}+\sin{\theta} \left(\beta_X ^2 \left(E_{+}^2-\frac{s}{2}\right)-s \tau_X\right)-\beta_X ^2 \left(2 E_{+}^2-s\right) \csc{\theta}\right] \Biggr\} \nonumber
\end{align}

\twocolumngrid

\subsection{Axion}
The matrix amplitude of the axion production has two t-channel contributions together with a third s-channel contribution coming from the photon coupling.
\begin{align}
    \mathcal{M}_1 =& \frac{e \, c_e}{f \left( t - m_e^2 \right)} \epsilon_\beta^{*} (q_1,\lambda_\gamma) \\
    &\Bar{v}(p_2,\lambda_{e^+}) \gamma^5 (\cancel{p_1} - \cancel{q_1} + m_e) \gamma^\beta u(p_1,\lambda_{e^-})\,, \nonumber \\
    \mathcal{M}_2 =& \frac{e \, c_e}{f \left(u - m_e^2 \right)} \epsilon_\beta^{*} (q_1,\lambda_\gamma) \\
    &\Bar{v}(p_2,\lambda_{e^+}) \gamma^\beta (\cancel{p_1} - \cancel{q_2} + m_e) \gamma^5 u(p_1,\lambda_{e^-})\,, \nonumber \\
    \mathcal{M}_3 =& \frac{e \, c_{\gamma}}{f s} g_{\mu \nu} \epsilon_\beta^{*} (q_1,\lambda_\gamma) \\
    &\Bar{v}(p_2,\lambda_{e^+}) \gamma^\mu u(p_1,\lambda_{e^-}) \epsilon^{\nu \beta \rho \sigma} (p_2 + p_1)_\rho (q_1)_\sigma\,. \nonumber
\end{align}
The axion coupling to fermions $c_e$ is proportional to $m_e$, hence its contribution will be proportional to $m_e^2$, and the interference between the two channels is suppressed by a factor of $m_e$ and is negligible. 
Following the same procedure as the hidden photon calculations, the amplitude is derived accounting for contributions from axion coupling to electrons, eq.~\eqref{eq:ALPb_amp}, axion coupling to photons, eq.~\eqref{eq:ALPs_amp}, and the interference between them, eq.~\eqref{eq:ALPint_amp}.
As described earlier, the two axion couplings have different dependencies on the axion mass.
The s-channel contribution is directly proportional to $(m_a^2-s)^2$, which for $m_a^2 \ll s$ simplifies to $s^2$, and as $m_a$ increases the expression tends to zero.
The t-channel has a more complicated dependence, where for small axion masses an overall suppression by $1/s$ emerges, and the s-channel dominates over the t-channel.
But as the mass increases and $m_a \sim \sqrt{s}$, the expression blows up and the t-channel starts to dominate.

\onecolumngrid
\begin{align}
    |\mathcal{M}|^2_{\text{t}} = &\frac{c_e^2 e^2 \csc^2{\theta}}{2 f^2 s (\tau_X-1)^2} 
    (E_{-} \cos{\theta}+E_{+})^2 \biggl[(\lambda_{e^-} \lambda_{e^+}+1) (1+ \tau_X^2) +\lambda_\gamma (\lambda_{e^-}+\lambda_{e^+}) \left(\tau_X^2-1\right)  \biggr]\,, \label{eq:ALPb_amp} \\
    |\mathcal{M}|^2_{\text{s}} = &\frac{c_\gamma^2 e^2 s (\tau_X-1)^2 }{32 f^2 (E_{-} \cos{\theta}+E_{+})^2}
    \Biggl\{ 2 \lambda_\gamma  (\lambda_{e^-}-\lambda_{e^+}) \left[ 4 E_2^2 \cos{\theta} + \left(E_{+}^2-\frac{s}{2}\right) (\cos{\theta}+1)^2\right] \label{eq:ALPs_amp} \\
    &\hspace{+5cm} + (\lambda_{e^-} \lambda_{e^+}-1) \left[ 4 E_2^2 \cos{\theta} - \left(E_{+}^2-\frac{s}{2} \right) (\cos{\theta}+1)^2 \right] \Biggr\}\,,  \nonumber \\
    |\mathcal{M}|^2_{\text{i}} = &\frac{c_\gamma c_e e^2 m_e \csc{\theta}}{2 f^2 s^2} 
     \Biggl\{ \lambda_\gamma \left[
     s (\lambda_{e^-}-\lambda_{e^+}) \left(E_{+}^2-\frac{s}{2}\right) (\cos{\theta}+1)^2 (1-\tau_X) - \frac{1}{2} s^2 (\lambda_{e^-}+\lambda_{e^+}) (\tau_X+1) \sin ^2(\theta ) \right] \label{eq:ALPint_amp} \\
     &\hspace{+2.5cm} + s (\tau_X-1) \left[ \lambda_{e^-} \lambda_{e^+} (E_{-}+E_{+} \cos{\theta})^2-(E_{-} \cos{\theta}+E_{+})^2\right]-E_2^2 \left[\cos (2 \theta )+3 \right]
     \Biggr\}\,. \nonumber
\end{align}

\twocolumngrid
The leading order terms for the contribution from the axion coupling to photons require the incoming fermions to have opposite helicity, and any contributions from equal helicities are of $\mathcal{O}(m_e^2)$.
Unlike the hidden photon and s-channel contribution, the t-channel contribution from the axion coupling to fermions requires equal beam helicities whereas opposite helicities only give rise to terms $\mathcal{O}(m_e^4)$.
Hence with $100\%$ polarised beams one can distinguish, to leading order approximation, between the two axion couplings. 
The s-channel contribution gives $\mathcal{R}\left(P_{e^-},P_{e^+} \right) = (1 - P_{e^-}P_{e^+})$  at $\mathcal{O}(m_e^0)$, while the t-channel contribution gives rise to $\mathcal{R}\left(P_{e^-},P_{e^+} \right) = (1 + P_{e^-}P_{e^+})$ at $\mathcal{O}(m_e^2)$, both with little to no corrections from higher orders.
The full axion amplitude has a more complex structure with the two main components, from each axion coupling, having opposite polarization dependencies. 
As axion coupling to photons dominates for small masses, for equal couplings, the maximum enhancement is found for opposite helicities, whereas the opposite becomes true for large mass.

For the axion coupling to photons the photon polarisation is 
\begin{align}
    \alpha_\gamma =  \frac{\left(P_{e^-}-P_{e^+} \right)}{\left(P_{e^-}P_{e^+}-1\right)}
     f\left(E_1,E_2\right)\,,
\end{align}
which, unlike the case of the hidden photon expression~\eqref{eq:a_g_DP}, doesn't dependent on the axion mass as implied by the helicity fraction distribution. Values of $|\alpha_\gamma| \approx +1$ are achieved for all $m_a$ when applying the forward angular cut $( 10^\circ \le \theta \le 50^\circ)$.
For the axion coupling to electrons,
\begin{align}
    \alpha_\gamma = \frac{\left(P_{e^-}+P _{e^+}\right) \left(s^2-m_a^4\right)}
    {\left(P_{e^-}P_{e^+}+1\right) \left(m_a^4+s^2\right)}\,,
\end{align}
 and the photon polarization is independent of $\cos{\theta}$, and maximum magnitude is found for small $m_a$ and fully polarised beams. $|\alpha_\gamma| \approx +1$ for $m_a \le 2$ GeV with no angular cuts applied.
The full axion contribution containing both axion coupling to electron and photons has elements from both, having $\left(P_{e^-} \pm P_{e^+} \right)$ term in the numerator and $\left(P_{e^-}P_{e^+} \pm 1\right)$ in the denominator. For opposite helicities the $m_a$ dependence disappears, whereas for equal helicities the angular dependence goes away, and we recover the contribution from each coupling as expected.

\onecolumngrid

\section{Standard Model Background} \label{app:SM_backs}
In the following, we display the individual $\theta_{\text{lab}} - E_{\text{CMS}}$ distributions for the four distinct contributions to the SM background considered.
Each panel has an individual color coding.

\begin{figure*}[h]
    \begin{subfigure}[b]{0.49\textwidth}
         \centering
        \includegraphics[width=\textwidth]{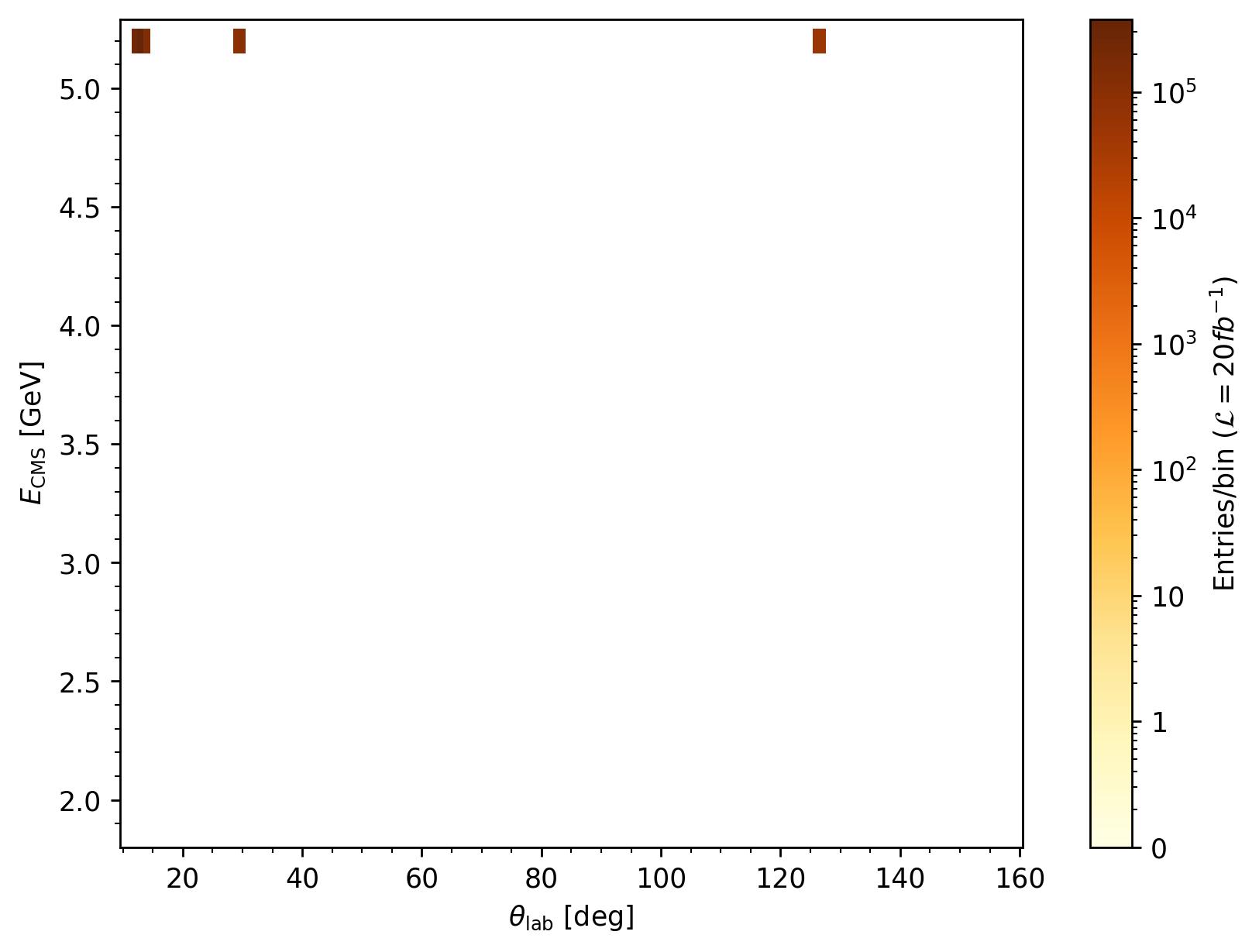}
    \end{subfigure}
    \hfill
    \begin{subfigure}[b]{0.49\textwidth}
         \centering
         \includegraphics[width=\textwidth]{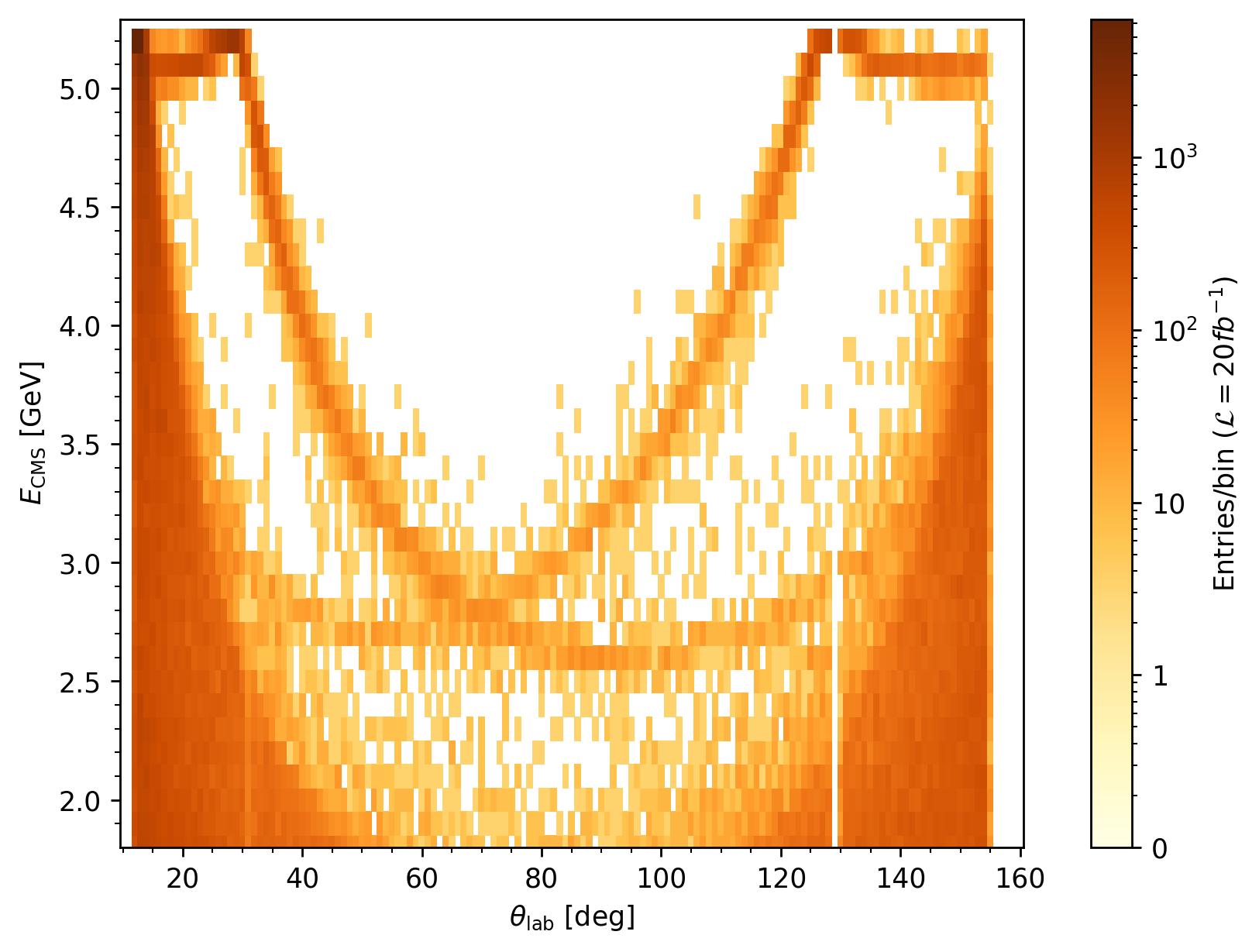}
    \end{subfigure}
    \hfill
    \begin{subfigure}[b]{0.49\textwidth}
         \centering
         \includegraphics[width=\textwidth]{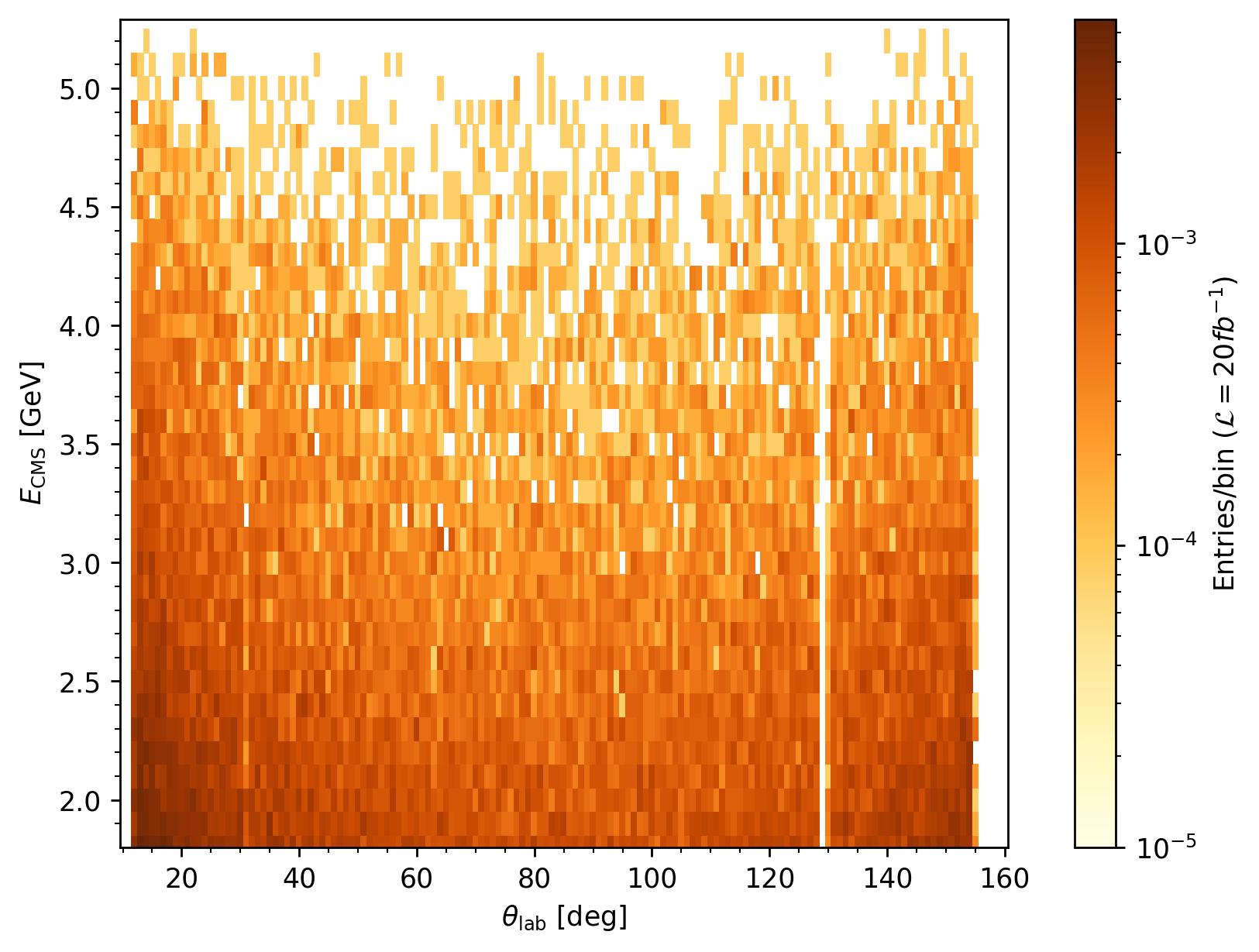}
    \end{subfigure}
    \hfill
    \begin{subfigure}[b]{0.49\textwidth}
         \centering
         \includegraphics[width=\textwidth]{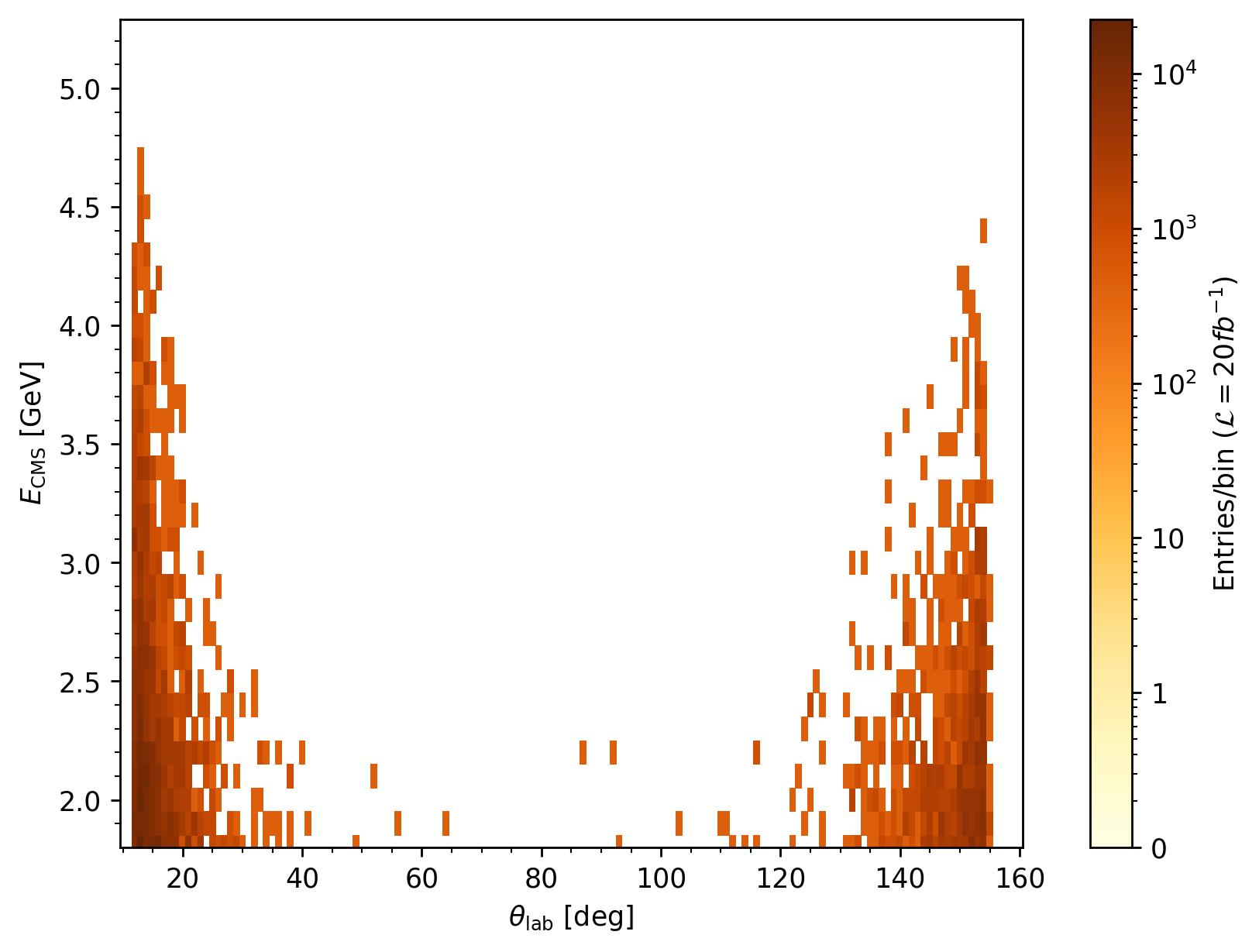}
    \end{subfigure}
    \caption{SM background distributions for $e^+ e^- \to$ $\gamma \gamma$ (upper left), $\gamma \gamma \gamma$ (upper right), $\nu \Bar{\nu} \gamma (\gamma)$ (bottom left), and $e^+ e^- \gamma (\gamma)$ (bottom right)}
    \label{fig:SM_back_2}
\end{figure*}

\clearpage
\twocolumngrid

\section{Exclusion Limits} \label{app:ex_lims}

\begin{figure}[h]
    \centering
    \begin{subfigure}[b]{0.49\textwidth}
        \includegraphics[width=\textwidth]{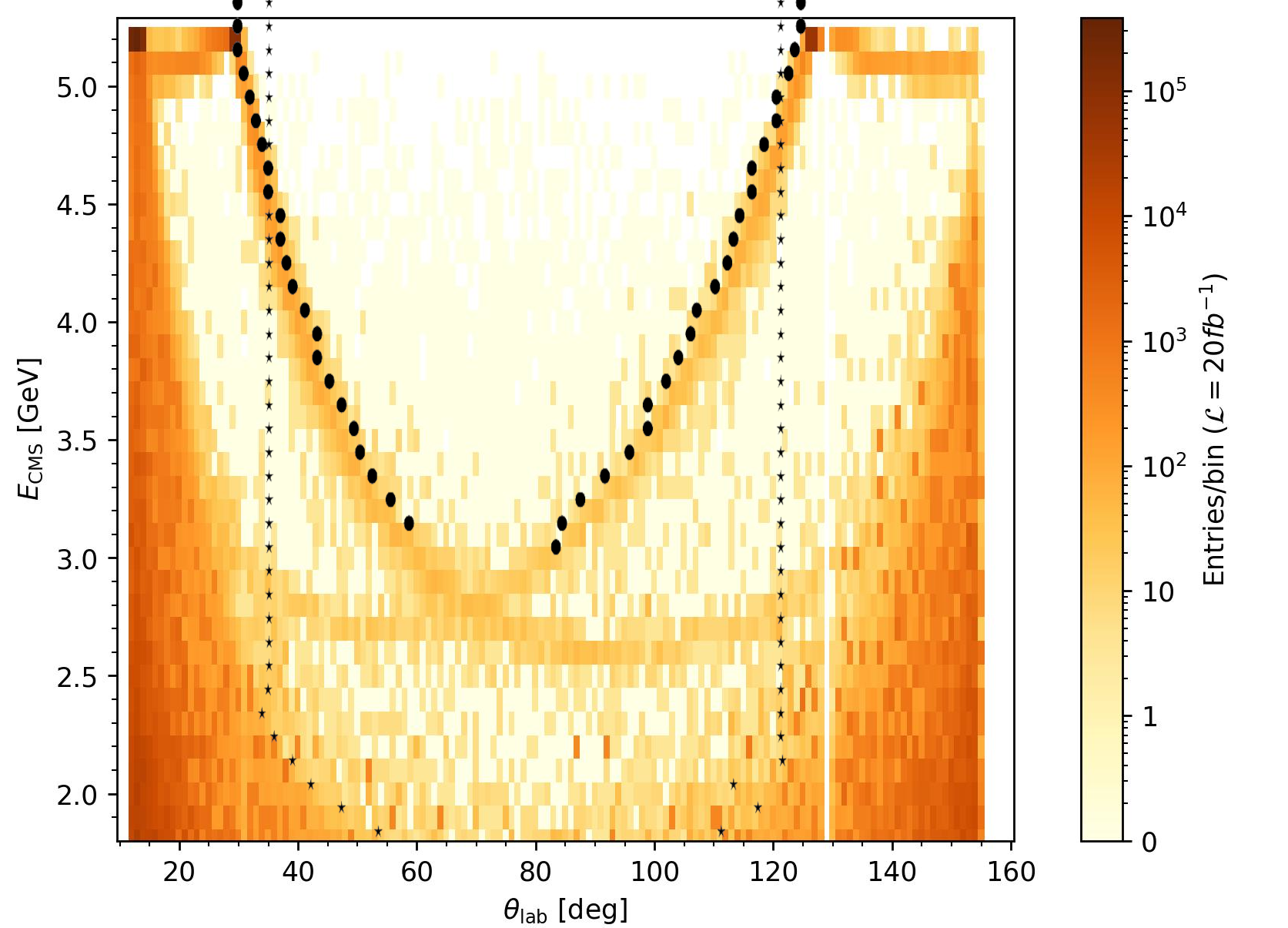}
    \end{subfigure}
    \begin{subfigure}[b]{0.49\textwidth}
        \includegraphics[width=\textwidth]{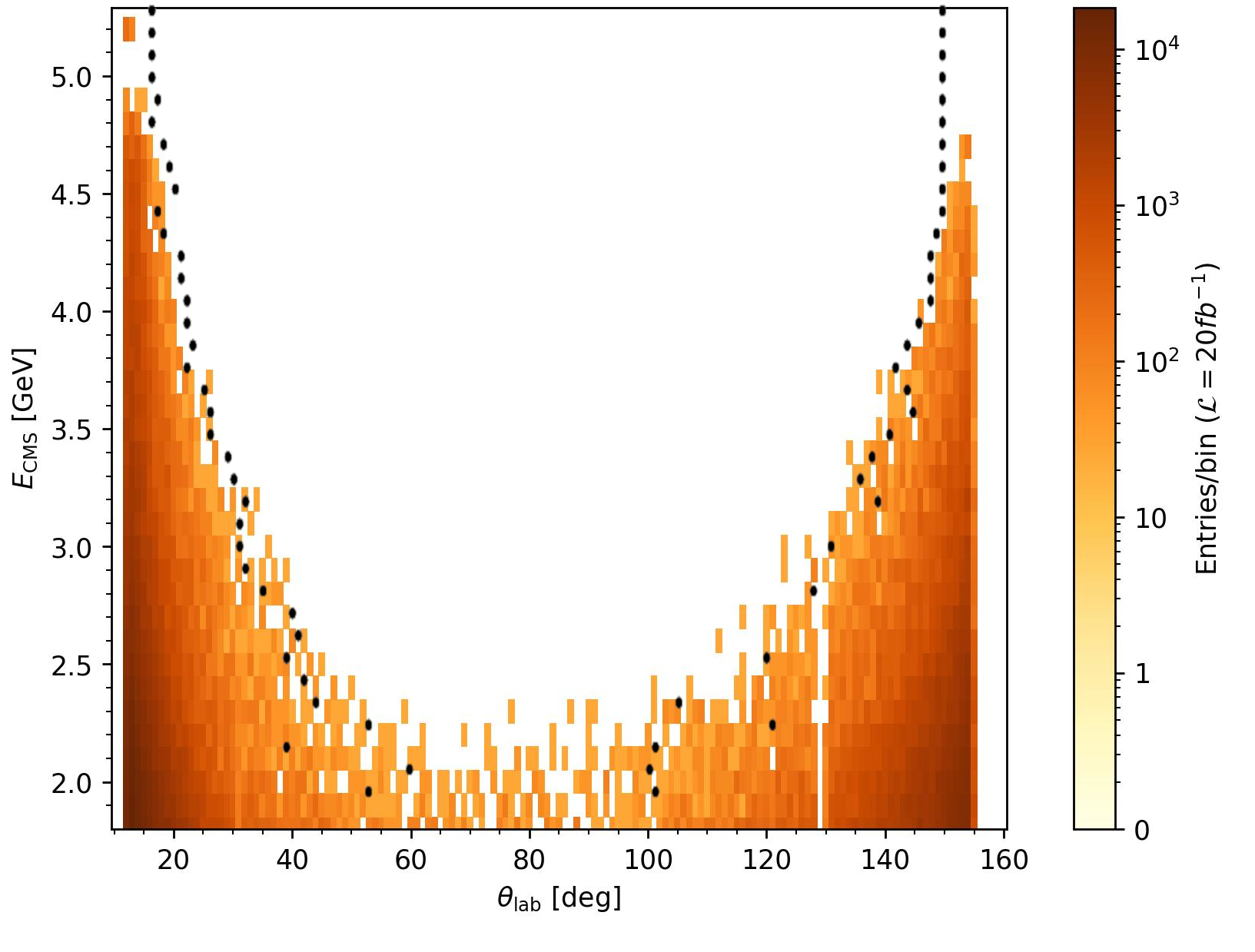}
    \end{subfigure}
    \caption{Points used for unpolarized (left) with low mass (black circles) and high mass (black stars),
    and equal beam polarization (right) background.}
    \label{fig:plot_points}
\end{figure}

In order to optimise the signal to background ratio, fits were performed, creating $\theta_{\text{lab}}-E_{\text{CMS}}$ cuts with the aim of selecting the regions with the least number of background events. In the following, we will present the work carried out to  perform the fits.
When observing the background in Figure~\ref{fig:SM_background}, two prominent regions stand out; the V-shape created by the $\gamma \gamma \gamma$ final state bands, and the areas between the main $e^+ e^- \gamma (\gamma)$ background and the bands.
The signal for low mass invisible states falls within the first region, but for increasing masses the second region has to be considered.
Two fits are therefore performed.
For the high mass fit, the points were found for each photon energy; going from small and large $\theta_{\text{lab}}$-values respectively, finding the first bin with less than 10 events.
A similar procedure was done for the low mass fit, this time going from the middle $(\theta_{\text{lab}} \approx 80^\circ)$ for both decreasing and increasing $\theta_{\text{lab}}$, the last bin with fewer events than our threshold were found. 
Furthermore, the three peaks from the $\gamma \gamma$ final state were of particular interest to avoid. 
This was ensured by finding the coordinates of the two centre peaks and continuing these points downwards (constant  $\theta_{\text{lab}}$ but decreasing $E_{\text{CMS}}$) until reaching the points found as described above.
As the outgoing photon in the case of large masses has very low energy, the fit was extended downwards into the area of increased background by including the range of $[60,100]^\circ$.
The points found are displayed on of the relevant distributions in Figure~\ref{fig:plot_points}.
The fits were found using Mathematica \cite{Mathematica:2023} for generic polynomial,
\begin{equation}
    c_1 + c_2 \, x + c_3 \, x^2 + c_4 \, \sqrt{x} + c_5 \, x^{-1/2} + c_6 \, x^{-1} + c_7 \, x^{-2} \,. \nonumber 
\end{equation}

The cut functions for the unpolarised SM background are described by eq.~\eqref{eq:E_CMS_low} and \eqref{eq:E_CMS_high} for low and high $m_X$ respectively.
For the equal beam helicity, we saw that the photon-only final states were not present, and therefore this opens a big area with little to no background (only from neutrino final states).
This allows for a single, much wider fit which include more signal events.
For equal beam polarization, the cut function can be found in eq.~\eqref{eq:E_CMS_eq}.

\onecolumngrid
\begin{align}
    E_{\text{CMS, low}} (\theta_{lab}) &= -1.753\times 10^4 - 61.42 \, \theta_{lab} + 4.708\times 10^{-2} \, \theta_{lab}^2 \label{eq:E_CMS_low} \\
    & \qquad +1.572\times 10^3 \sqrt{\theta_{lab}}+\frac{9.795\times 10^4}{\sqrt{\theta_{lab}}}-\frac{2.379\times 10^5}{\theta_{lab}}+\frac{6.994\times 10^5}{\theta_{lab}^2} \,, \nonumber \\
    E_{\text{CMS, high}} (\theta_{lab}) &= - 2.601\times 10^4 - 1.109\times 10^2 \, \theta_{lab} + 0.1001 \, \theta_{lab}^2 \label{eq:E_CMS_high} \\
    & \qquad +2.583\times 10^3 \sqrt{\theta_{lab}} + \frac{1.304\times 10^5}{\sqrt{\theta_{lab}}} - \frac{2.826\times 10^5}{\theta_{lab}} + \frac{6.573\times 10^5}{\theta_{lab}^2} \,, \nonumber \\ 
    E_{\text{CMS, equal}}(\theta_{lab}) &= - 2.929\times 10^3 - 13.66 \, \theta_{lab} + 1.293\times 10^{-2} \, \theta_{lab}^2 \label{eq:E_CMS_eq} \\
    & \qquad + 3.061\times 10^2 \sqrt{\theta_{lab}} + \frac{1.379\times 10^4}{\sqrt{\theta_{lab}}} - \frac{2.763\times 10^4}{\theta_{lab}} + \frac{5.284\times 10^4}{\theta_{lab}^2} \,. \nonumber
\end{align}

\clearpage

\twocolumngrid

\section{Helicity Fraction Distributions} \label{app:hel_dia}
In the following, we present explanations with visual aid for the helicity fraction distributions shown in this paper.
The observable considered is the fraction of the final state photon with helicity matching that of the incoming electron beam as a function of the angle of the photon,
\begin{equation}
    P = \frac{|\mathcal{M}|^2 (2\lambda_{e^-} = \lambda_\gamma = 1)}{|\mathcal{M}|^2 (\lambda_{e^\pm} = \lambda_\gamma = 0)}\,,
\end{equation}

where $|\lambda_\gamma| = 1$ is the helicity of the photon, and $|\lambda_{e^-}| = 1/2$ the helicity of the electron.
Hence at each $\cos{\theta_{\text{lab}}}$-value, the distribution indicates that for $X$ photons detected at this angle $P \times X$ of them will have the same helicity as the incoming electron beam, and $(P-1) \times X$ will have the opposite helicity.

The representations below are 2D diagrams with a horizontal z-axis and arbitrary vertical axis in the x-y plane. 
Without loss of generality, the electron is assumed to have positive helicity,
and two combinations of incoming beam helicities are possible; 
the electron and positron spin vectors are parallel ($\lambda_{e^+} = -\lambda_{e^-}$) or antiparallel ($\lambda_{e^+} = \lambda_{e^-}$).

\begin{figure}[b]
     \centering
     \includegraphics[width=0.49\textwidth]{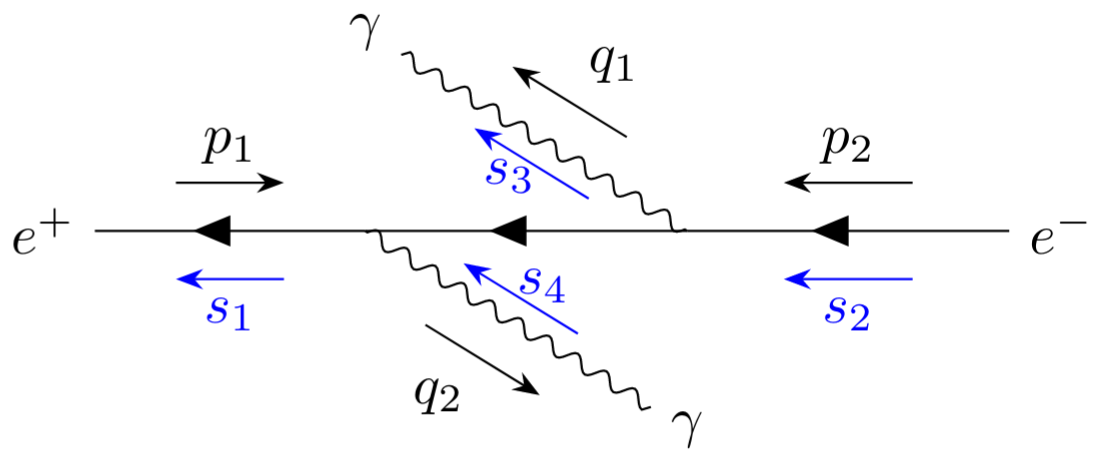}
     \caption{Helicity visualisation for $e^+ e^- \rightarrow \gamma \gamma$}
     \label{fig:spin_ee_aa}
\end{figure}

\subsection{Standard Model Background}\label{app:SM_hel}
\begin{figure}[t]
    \centering
    \includegraphics[width=0.49\textwidth,trim={0 0 0 1cm},clip]{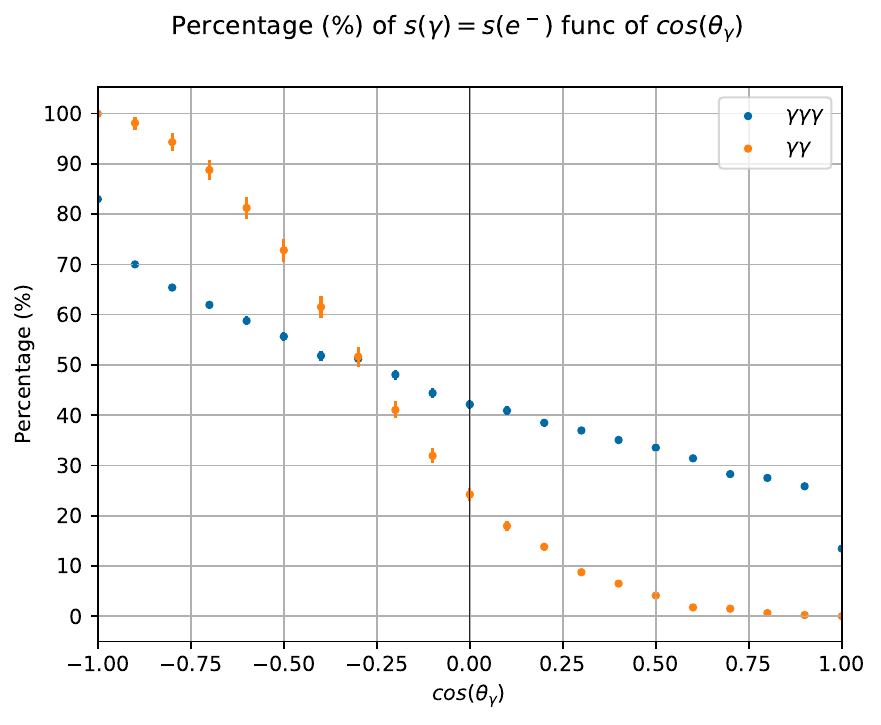}
    \caption{Helicity fraction distributions for $e^+ e^- \rightarrow \gamma \gamma (\gamma)$ with no cuts.}
    \label{fig:SMa_hel_cos}
\end{figure}

The helicity fraction distribution for the background has contributions from every SM background process, each with their own tendencies which will be described in this section.
The simplest case is $e^+ e^- \rightarrow \gamma \gamma$, where the incoming electron and positron must have parallel spin vectors (opposite helicities) due to the emission of two photons not changing change the direction of the spin (see Figure~\ref{fig:spin_ee_aa}).
The $z$-components of the spin vectors of the outgoing photons will have the same sign as the incoming fermions.
The direction of the photon will determine its helicity; if emitted in the same direction as the electron, the helicity will match that of the electron, and vice versa if emitted in the positron direction.
Hence the helicity fraction distribution has the s-shape seen on Figure~\ref{fig:SMa_hel_cos} due to the photon being more likely to travel in the same direction as the fermion which it was emitted from.
 At $\cos{\theta_{\text{lab}}} = -1$, all photons detected will have the same helicity as the incoming electron, whereas in the opposite direction, $\cos{\theta_{\text{lab}}} = +1$,  none of them will and their helicity match the incoming positron.
The skew away from the center line is due to the unequal beam energies at Belle II, for equal energies the shape will be centered around $(0,50 \%)$.

The spin vector diagram looks similar for $e^+ e^- \rightarrow \gamma \gamma \gamma$, the fermions are still required to have opposite helicities, but the helicities of the outgoing photons are more random due to the 3-particle final state. Therefore the correlation between direction and helicity fraction reduces, and the distribution is closer to $50 \%$. 
Figure \ref{fig:SMa_hel_cos} displays the helicity fraction distribution for $e^+ e^- \rightarrow \gamma \gamma (\gamma)$ without any cuts applied.
When implementing the criteria that only one photon can be detected, the few cases allowed for $e^+ e^- \rightarrow \gamma \gamma$ are due to the asymmetric angular coverage of the detector and the gaps between the barrel and end-caps. As seen on the lower-right of the left panel on Figure~\ref{fig:hel_cos}, this corresponds to the two points around $\cos{\theta} \approx -1$ and another set at $\cos{\theta} \approx 0.6$.

The helicities involved in the main background process, $e^+ e^- \rightarrow e^+ e^- \gamma (\gamma)$, are much more complex as consists of many contributions and interferences between them, some of which require the incoming helicities to be opposite and others equal.
The resulting angular distribution, upper-left on the left panel of Figure~\ref{fig:hel_cos}, does not show any directional dependence. 
Similar arguments can be made for $e^+ e^- \rightarrow \nu \Bar{\nu} \gamma (\gamma)$, and as seen on the upper-right of Figure~\ref{fig:hel_cos}, the process has a slight  $\cos{\theta}$-dependence, but mostly resides around $50 \%$.

As $e^+ e^- \rightarrow e^+ e^- \gamma$ is the dominating process, the full background results, displayed in the lower-left of Figure~\ref{fig:hel_cos}, looks very similar. The slight upwards motion at $\cos{\theta} \approx -1$ and outlying point around $\cos{\theta} \approx 0.6$ are due to the large $e^+ e^- \rightarrow \gamma \gamma$ contribution.

\begin{figure}[b]
    \begin{subfigure}[b]{0.45\textwidth}
         \centering
        \includegraphics[width=\textwidth]{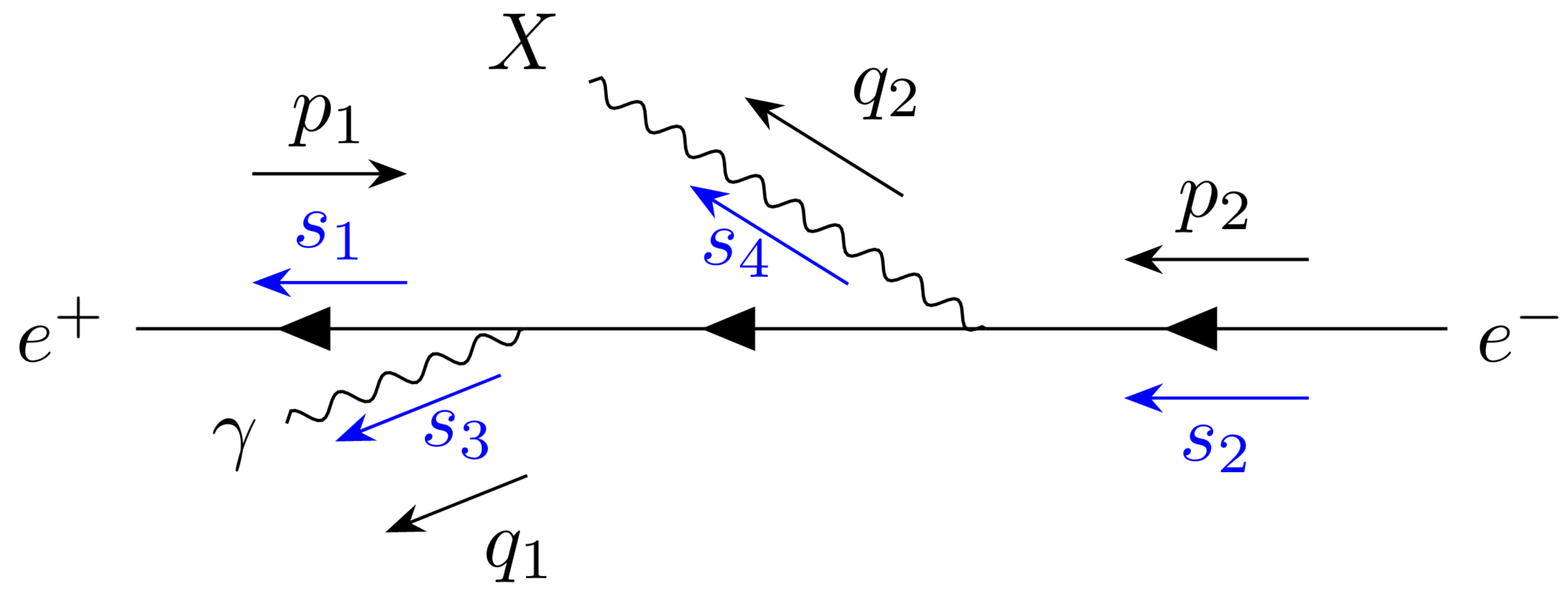}
    \end{subfigure}
    
    \begin{subfigure}[b]{0.45\textwidth}
         \centering
         \includegraphics[width=\textwidth]{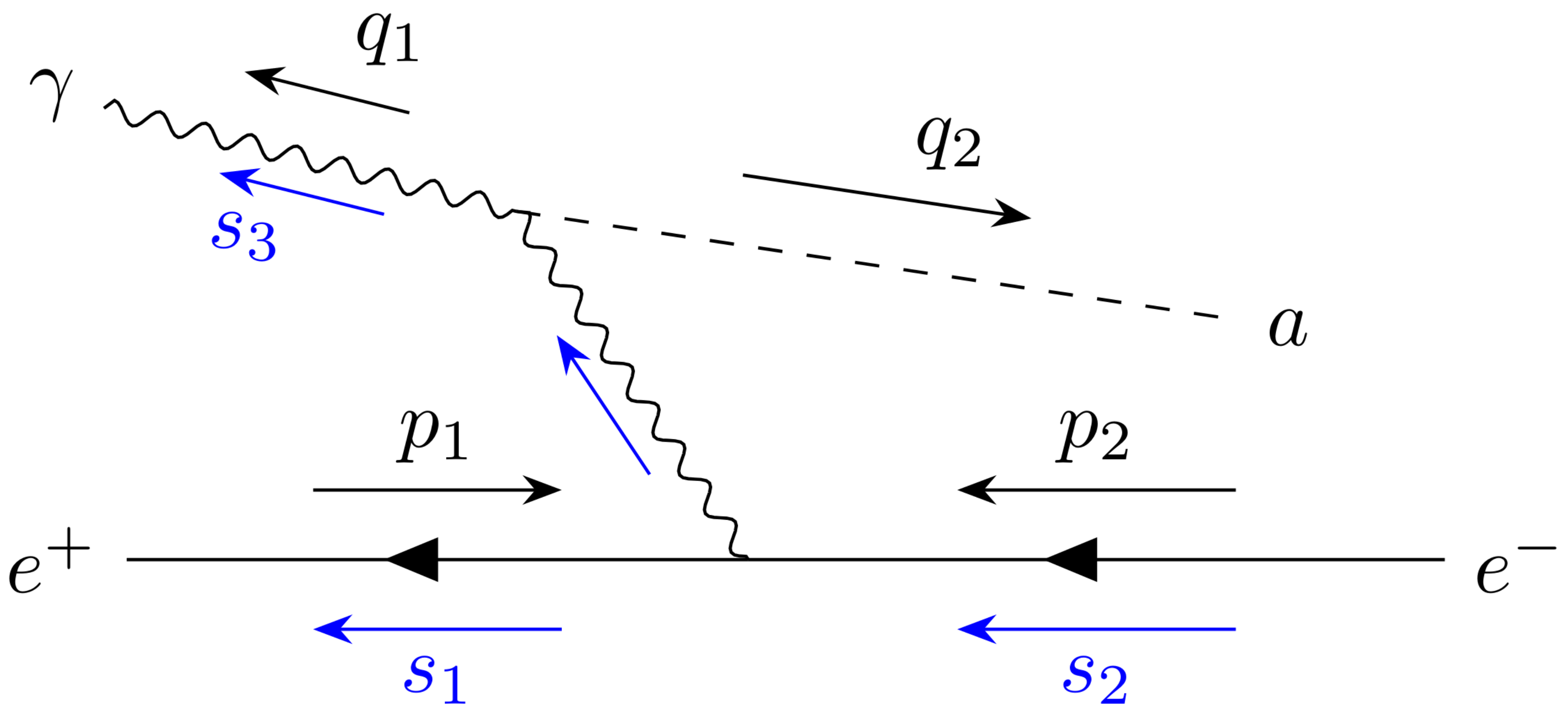}
    \end{subfigure}
    
    \begin{subfigure}[b]{0.45\textwidth}
         \centering
         \includegraphics[width=\textwidth]{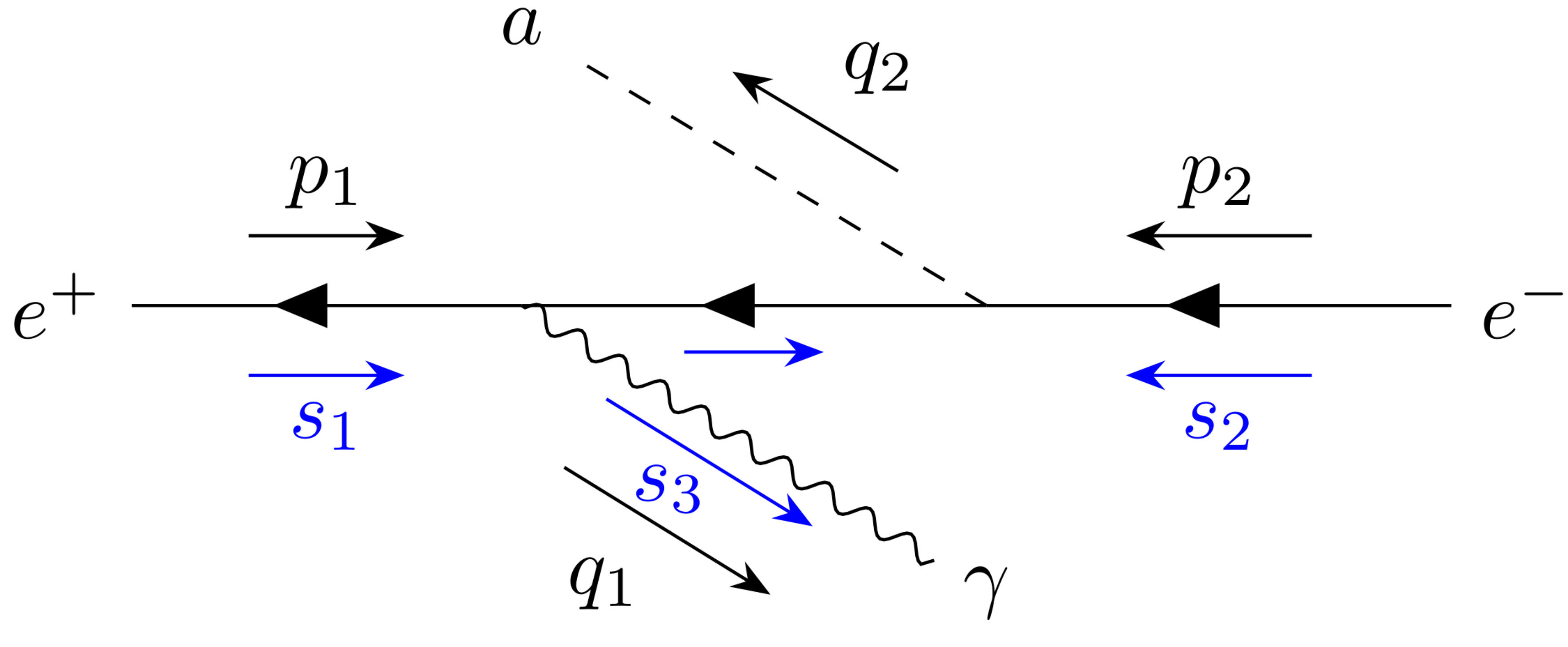}
    \end{subfigure}
    \caption{Helicity visualisation for the  production of hidden photons for large $m_X$ (top), axion-photon (middle) and axion-electrons (bottom) couplings.}
    \label{fig:spin_ALP}
\end{figure}

\subsection{Hidden Photon} \label{app:hel_DP}

The hidden photon interacts similarly to the SM photon, and hence the incoming fermions are required to have opposite helicities.
Two scenarios are possible; the photon and hidden photon travel in opposite directions along the $z$-axis (Figure~\ref{fig:spin_ee_aa} with one $\gamma$ replaced with $X$), or due to the unequal beam energies they travel in the same direction (see top panel on Figure~\ref{fig:spin_ALP}).
The direction of the spin vectors involved remain the same, but as the momentum changes the photon helicity differs between the two scenarios. 
For small $m_X$ effectively only scenario one happens, resulting in the photon direction and helicity being strongly coupled; the helicity of the photon will match the helicity of the fermion travelling in the same direction. 
But as the hidden photon mass increases, the photon becomes softer and its direction more random. This introduces occurrences of scenario two, pushing the helicity fraction distribution towards $50 \%$. The results for the hidden photon can be seen on the right panel on Figure~\ref{fig:hel_cos}.

\subsection{Axions} \label{app:hel_ALPs}
\begin{figure}[b]
    \begin{subfigure}[b]{0.49\textwidth}
         \centering
         \includegraphics[width=\textwidth]{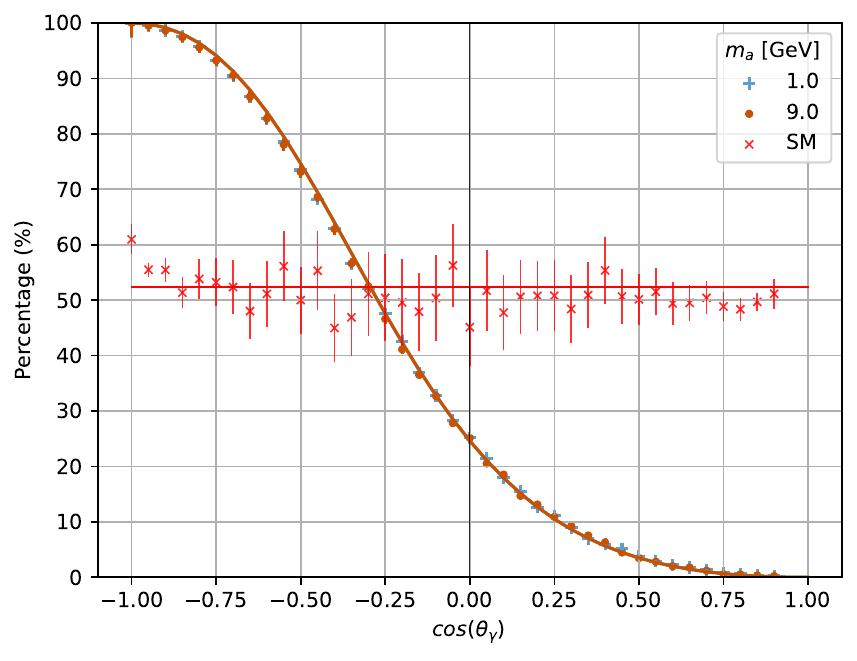}
    \end{subfigure}
    \begin{subfigure}[b]{0.49\textwidth}
         \centering
         \includegraphics[width=\textwidth]{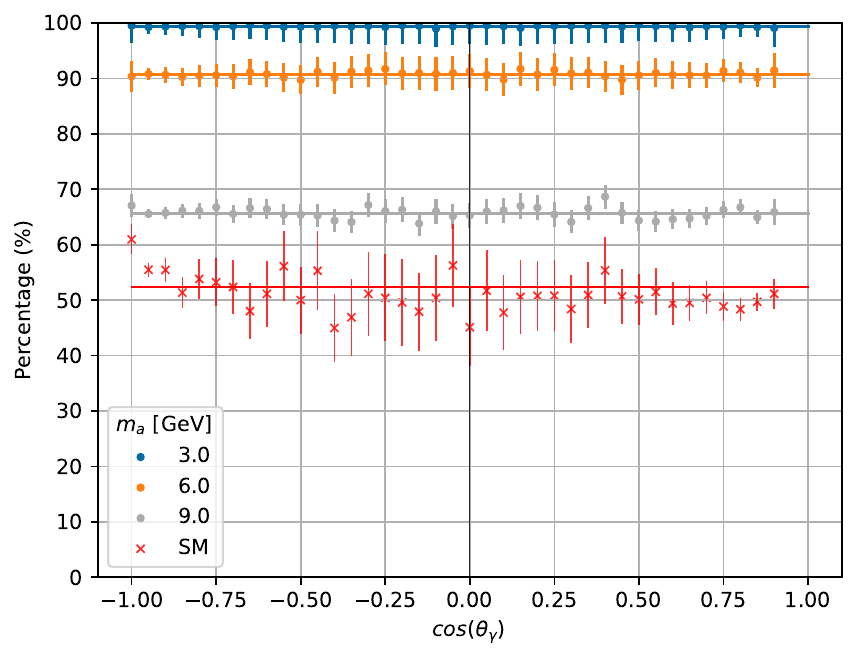}
    \end{subfigure}
    \caption{Helicity fraction distributions for axion-photon (top) and axion-electrons (bottom) couplings.}
    \label{fig:ALP_hel_cos_2}
\end{figure}

As the two axion contributions have different spin structures, they will be described separately. 
For axion coupling to photons, the incoming fermions annihilate into a virtual photon which then emits an axion.
The incoming fermions are required to have the same direction spin vector which the virtual photon inherits. 
The outgoing photon helicity is determined by its direction after the axion emission, see middle panel of Figure \ref{fig:spin_ALP}, and depends on whether it is in the same direction as the electron or positron.
Hence the helicity fraction distribution, top panel on Figure \ref{fig:ALP_hel_cos_2}, is the s-shape seen before.
The distribution is not influenced by the mass of the axion, as it only changes the coupling strength.
The axion coupling to electrons consists of t/u-channel diagrams like the hidden photon, but as the interaction introduces a helicity flip for the incoming fermions (bottom panel of Figure~\ref{fig:spin_ALP}), their helicities are equal. Therefore the photon helicity will match both fermions or neither, resulting in no angular dependence as seen on bottom panel on Figure \ref{fig:ALP_hel_cos_2}.
As described before, as the axion mass increases the photon direction becomes more random resulting in the helicity fraction approaching $50 \%$.

\clearpage

%%%%%%%%%%%%%%%%%%%%%%%%%%%%%%%%%
%%%%%%%%%%%%%%%%%%%%%%%%%%%%%%%%%

\end{document}